\documentclass[journal]{IEEEtran}
\ifCLASSINFOpdf
  \usepackage[pdftex]{graphicx}
  \graphicspath{{../pdf/}{../jpeg/}}
  \DeclareGraphicsExtensions{.pdf,.jpeg,.png,.bmp}
\else
  \usepackage[dvips]{graphicx}
  \graphicspath{{../jpg/}}
  \DeclareGraphicsExtensions{.jpg}
  \fi
\usepackage{url}
\usepackage{xurl}
\usepackage{graphicx}

\usepackage{balance}
\usepackage{makecell}
\usepackage{stfloats}
\usepackage{color,soul}
\usepackage{amsmath}
\usepackage{multirow}
\usepackage{hhline}
\usepackage{amssymb}
\usepackage{amsthm}
\usepackage{mathtools}
\usepackage{multicol}
\usepackage{comment}
\usepackage{adjustbox}
\usepackage{blindtext}
\usepackage{hyperref}
\definecolor{Gray}{gray}{0.9}
\usepackage{tikz}
\newcolumntype{M}[1]{>{\centering\arraybackslash}m{#1}}

\usepackage{xcolor,cite,etoolbox}
\makeatletter 
\pretocmd\@bibitem{\color{black}\csname keycolor#1\endcsname}{}{\fail}
\newcommand\citecolor[1]{\@namedef{keycolor#1}{}}
\makeatother
\citecolor{}

\usepackage{booktabs, makecell, tabularx}
\newcolumntype{C}{>{\centering\arraybackslash}X} 

\usepackage{stfloats}
\usepackage{siunitx}
\usepackage{caption}
\usepackage{xcolor,colortbl}
\usepackage{enumitem}

\hypersetup{
     colorlinks   = true,
     citecolor    = red,
     linkcolor    = red,
     urlcolor     = blue,
}

\usepackage{tikz}
\newcommand*\circled[1]{\tikz[baseline=(char.base)]{
            \node[shape=circle,draw,inner sep=0.5pt] (char) {#1};}}

\allowdisplaybreaks 

\begin{document}
\bstctlcite{IEEEexample:BSTcontrol}

\title{\huge{Integration of TinyML and LargeML: \\A Survey of 6G and Beyond}}

\author{Thai-Hoc~Vu, Ngo Hoang Tu,~\IEEEmembership{Member,~IEEE}, Thien Huynh-The,~\IEEEmembership{Senior Member,~IEEE},\\
Miroslav Voznak,~\IEEEmembership{Senior Member,~IEEE}, Kyungchun Lee,~\IEEEmembership{Senior Member,~IEEE},\\
Sunghwan Kim,~\IEEEmembership{Senior Member,~IEEE}, and Quoc-Viet~Pham,~\IEEEmembership{Senior Member,~IEEE} 

\thanks{The work of Miroslav Voznak is funded by the European Union under the REFRESH – Research Excellence For REgion Sustainability and High-tech Industries project number CZ.10.03.01\/00\/22\_003\/0000048 via the Operational Programme Just Transition, and also supported by the Ministry of Education, Youth and Sports of the Czech Republic (MEYS CZ) within a Student Grant Competition in the VSB – Technical University of Ostrava under project ID No. SGS SP2026\/004. The work of Kyungchun Lee is supported by the Basic Science Research Program through the National Research Foundation of Korea (NRF) funded by the Ministry of Education under Grant NRF-2019R1A6A1A03032119, and the NRF Grant funded by the Korean Government (MSIT) under Grant NRF-2022R1A2C1006566. The work of Sunghwan Kim is supported by the Research Program through the National Research Foundation of Korea under Grant NRF-2023R1A2C1003546. The work of Quoc Viet Pham is supported in part by Research Ireland under the European Innovation Council CHIST-ERA SHIELD project (Project No. 216449, Award No. 19226).}

\thanks{Thai-Hoc Vu and Miroslav Voznak (\textit{co-first author}) are with the Faculty of Electrical Engineering and Computer Science, VSB-Technical University of Ostrava, 17. Listopadu 2172/15, 708 00, Ostrava, Czechia (e-mail: thai.hoc.vu@vsb.cz, miroslav.voznak@vsb.cz).
Ngo~Hoang~Tu is with the Faculty of Information Technology, Van Lang School of Technology, Van Lang University, Ho Chi Minh City 70000, Vietnam (e-mail: tu.nh@vlu.edu.vn).
Thien~Huynh-The is with the Department of Electronics and Information Engineering, Ho Chi Minh City University of Technology and Engineering, Vietnam (e-mail: thienht@hcmute.edu.vn).
Kyungchun~Lee is with the Department of Electrical and Information Engineering and the Research Center for Electrical and Information Technology, Seoul National University of Science and Technology, Seoul 01811, South Korea (e-mail: kclee@seoultech.ac.kr).
Sunghwan Kim (\textit{corresponding author}) is with the Department of Artificial Intelligence and Information Technology, Sejong University, Seoul 05006, Republic of Korea (email: ks@sejong.ac.kr).
Quoc-Viet~Pham is with the School of Computer Science and Statistics, Trinity College Dublin, Dublin 2, D02 PN40, Ireland (e-mail: viet.pham@tcd.ie).}
}

\maketitle

\begin{abstract}
The evolution from fifth-generation (5G) to sixth-generation (6G) networks is driving an unprecedented demand for advanced machine learning (ML) solutions. Deep learning has already demonstrated significant impact across mobile networking and communication systems, enabling intelligent services such as smart healthcare, smart grids, autonomous vehicles, aerial platforms, digital twins, and the metaverse. At the same time, the rapid proliferation of resource‑constrained Internet‑of‑Things (IoT) devices has accelerated the adoption of tiny machine learning (TinyML) for efficient on‑device intelligence, while large machine learning (LargeML) models continue to require substantial computational resources to support large‑scale IoT services and ML‑generated content. These trends highlight the need for a unified framework that integrates TinyML and LargeML to achieve seamless connectivity, scalable intelligence, and efficient resource management in future 6G systems.
This survey provides a comprehensive review of recent advances enabling the integration of TinyML and LargeML in next‑generation wireless networks. In particular, we (\textit{i}) provide an overview of TinyML and LargeML, (\textit{ii}) analyze the motivations and requirements for unifying these paradigms within the 6G context, (\textit{iii}) examine efficient bidirectional integration approaches, (\textit{iv}) review state‑of‑the‑art solutions and their applicability to emerging 6G services, and (\textit{v}) identify key challenges related to performance optimization, deployment feasibility, resource orchestration, and security. Finally, we outline promising research directions to guide the holistic integration of TinyML and LargeML for intelligent, scalable, and energy‑efficient 6G networks and beyond.
\end{abstract}


\IEEEpeerreviewmaketitle

\section{Introduction}
\label{sec:Introduction}
\begin{figure*}[!t]
\centering
\includegraphics[width=.88\linewidth]{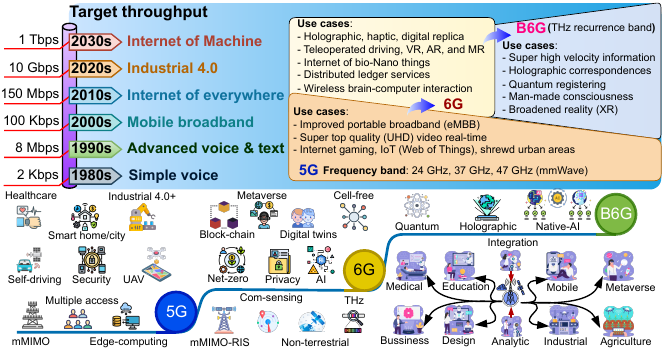}
\caption{Transition from 5G to 6G and beyond 6G (B6G): Target throughput, use cases, core technologies, and applications.}
\label{Fig:transition}
\end{figure*}

The evolution of communication technologies has been characterized by extraordinary advances every decade. From the first-generation (1G) analog systems, which introduced mobile voice communications, to {fifth-generation (5G)} networks currently transforming industries with ultra-fast speeds and low latency, each generation has delivered major transformative impacts \cite{dao2024review}. Looking ahead, {sixth-generation (6G)} networks are expected to push technological boundaries further, offering unprecedented capabilities and driving global innovation. 
The core features of 6G will clearly differentiate it from previous generations (Fig.~\ref{Fig:transition}), positioning the paradigm as a transformative technology for diverse applications. These features address the escalating demands for higher data rates, superior reliability, efficient spectrum management, and superintelligence capabilities. Furthermore, 6G aims to seamlessly integrate advanced artificial intelligence (AI) and machine learning (ML) technologies, expanding the possibilities of wireless networks and significantly improving the delivery of high-end applications and services to users. 

The integration of AI and ML into 6G will enable new levels of automation and intelligence, unlocking numerous new applications and high-end services \cite{zhang2020towards}. AI already operates on multiple network layers to improve dynamic resource management, predict maintenance needs, and personalize user experiences. Advances in ML, particularly deep learning (DL) architectures that execute computer vision and natural language processing (NLP) tasks, are expected to support high-fidelity holographic communications. This capability will allow users to interact in immersive three-dimensional environments and could revolutionize remote collaboration. In a hyper-connected future, 6G will unlock the full potential of the Internet-of-Things (IoT) and mobile big data to drive innovation and lead the world into smarter, deeper connectivity.

\subsection{Context and Motivation}
\begin{table*}[ht!]
    \centering
    \caption{Comparison of TinyML and LargeML Features.}
    \begin{tabular}{|m{3.5cm}|m{6.4cm}|m{6.8cm}|}
        \hline
        \multicolumn{1}{|c|}{\textbf{Aspects}} & \multicolumn{1}{c|}{\textbf{TinyML}} & \multicolumn{1}{c|}{\textbf{LargeML}} \\
        \hline\hline
        \textbf{Definition} & ML on resource-constrained devices & Large-scale AI models with vast numbers of parameters and large data processing capacity \\
        \hline
        \textbf{Computational requirements} & Limited power,  memory, and processing capability & Substantial power,  memory, and computational resources \\
        \hline
        \textbf{Deployment environment} & Smart sensors, edge devices, and microcontrollers & Cloud servers and specialized hardware (e.g., GPUs, TPUs) \\
        \hline
        \textbf{Latency} & Ultra-low latency for real-time processing & Higher latency, acceptable for batch processing \\
        \hline
        \textbf{Power consumption} & Very low, suitable for battery-powered devices & High, requiring a robust power supply \\
        \hline
        \textbf{Data processing} & Local (on-device) processing & Centralized (large-scale) data processing \\
        \hline
        \textbf{Typical applications} & Real-time analytics and sensor data processing in IoT environments (e.g., smart homes, industrial monitoring) & Complex data analysis and decision-making (e.g., NLP, image recognition, autonomous systems) \\
        \hline
        \textbf{Advantages} & Low latency, reduced data transmission, enhanced privacy & High accuracy, with the ability to process and analyze large datasets \\
        \hline
        \textbf{Challenges} & Constrained by hardware limitations and task complexity & Requires extensive computational infrastructure and incurs high costs \\
        \hline
        \textbf{Complementary roles} & Supports real-time, lightweight, and edge-based tasks & Enables deep, comprehensive analysis and cloud-based decision-making \\
        \hline
        \textbf{Examples} & Keyword spotting, anomaly detection in sensor data & GPT-4, BERT, large-scale image classifiers \\
        \hline

        \hline
        
    \end{tabular}
    \label{tab_tinyml_vs_large_models}
\end{table*}

In the last decade, AI has converged around two contrasting approaches: tiny machine learning (TinyML) and large-scale machine learning (LargeML). TinyML focuses on creating efficient models for resource-constrained edge devices, allowing real-time analytics close to data sources (i.e., bringing analytics closer to data sources) \cite{Lin2023Oct}. In contrast, LargeML leverages powerful cloud servers to handle complex, data-intensive tasks, exemplified by models like GPT-4 and BERT, which excel in applications such as sentiment analysis and creative text generation \cite{xu2024large} {(e.g., Large Language Models (LLMs), generative AI (GenAI), and Agentic AI)}.

{However, LargeML's heavy computational demands limit real‑time use on constrained devices and raise privacy concerns, while TinyML mitigates these issues by pre‑processing data at the edge and enabling LargeML to deploy compressed models through knowledge distillation \cite{zhu2021data}. This collaborative strategy improves flexibility for IoT services: TinyML gains accuracy from LargeML's advanced learning, and LargeML benefits from TinyML's real‑time processing and privacy preservation, as summarized in Table~\ref{tab_tinyml_vs_large_models}. Their integration strengthens AI capabilities in 6G systems, for example, wearable health monitors can analyze vital signs locally with TinyML \cite{tsoukas2021review} while LargeML detects medical risks from historical data, and TinyML can process Electroencephalography (EEG) signals for emotional state recognition \cite{nguyen2023behind} while LargeML translates them into metaverse actions \cite{wang2025large}.

Despite these advantages, challenges remain, including heightened security risks from massive device connectivity and ethical concerns around explainability and data privacy, yet combining TinyML and LargeML ultimately enables seamless device-server interaction and advances in data-driven innovation and intelligent living environments.}

\subsection{State-of-the-Art and Contributions}
\label{sec:Introduction_Contributions}

{ Over the past decade, numerous surveys have examined TinyML's role in edge computing, offering comprehensive overviews of its fundamentals, development tools, applications, and future directions \cite{dutta2021tinyml, abadade2023tinyml, ray2022review, rajapakse2023intelligence, tsoukas2024review, capogrosso2024machine}, with key contributions including hardware–software co‑design for efficient deployment on constrained devices \cite{dutta2021tinyml} and systematic reviews of TinyML toolchains spanning hardware platforms, software frameworks, and libraries \cite{ray2022review}. Other works \cite{abadade2023tinyml, tsoukas2024review, capogrosso2024machine} highlight TinyML's broad application landscape, from anomaly detection and predictive maintenance to intelligent sensor networks, where models monitor sensor data for deviations, forecast equipment failures, and enable fast, local decision‑making. Collectively, these surveys highlight TinyML's transformative potential across industrial automation, environmental monitoring, smart infrastructure, and other data‑driven domains.}

Despite these advances, the existing survey literature on TinyML reveals some limitations when considered in the context of 6G integration. Although these surveys examine TinyML applications within general IoT frameworks, they overlook the specific challenges and opportunities presented by integrating TinyML with LargeML in 6G networks~\cite{abadade2023tinyml}. 
Some surveys~\cite{tsoukas2024review, capogrosso2024machine} provide limited clarification on fine-tuning TinyML models to the specific resource constraints of 6G environments.
For example, 6G may require the development of lightweight communication protocols tailored to TinyML models for efficient data exchange with server-side LargeML systems. These surveys also fall short of exploring optimization strategies for minimizing power consumption in TinyML models while preserving the accuracy required for effective collaboration with LargeML~\cite{capogrosso2024machine}.

{In contrast, LargeML surveys outline complementary strengths and limitations, with works such as \cite{raiaan2024review, chang2024survey, han2024parameter, zhao2024explainability, min2023recent} examining architectures, applications, evaluation methods, and open challenges. Reviews like \cite{raiaan2024review} analyze transformer-based models and LLM development, while \cite{min2023recent} highlights the impact of large pre-trained language models (PLMs) on diverse NLP tasks, emphasizing their strong performance in text generation, translation, and sentiment analysis. Other surveys \cite{chang2024survey, han2024parameter, zhao2024explainability} focus on challenges such as explainability and the need for parameter-efficient fine-tuning (PEFT) \cite{han2024parameter} to reduce computational overhead. However, despite their breadth, current LargeML surveys do not address 6G‑specific constraints: they overlook deployment and coordination challenges in real‑time, latency‑sensitive, edge‑intelligent networks, and the need to compress or adapt LargeML models for efficient operation alongside TinyML systems. They also omit the implications of emerging 6G communication protocols and strict latency requirements for training and inference. Table~\ref{Table:Summary_ExistingSurveys} summarizes these surveys and situates the contributions of this work.}

\begin{table*}[ht!]
    \renewcommand{\arraystretch}{1}
	\caption{Summary of related surveys on TinyML and LargeML.}
	\label{Table:Summary_ExistingSurveys}
	\centering
    \resizebox{\linewidth}{!}{
	\begin{tabular}{|p{0.65cm}|c|c|p{7.75cm}|p{5.50cm}|}
		\hline 
		\multirow{2}{*}{\textbf{Refs.}}  & \multicolumn{2}{c|}{\textbf{Review Topics}}  & \multicolumn{1}{c|}{\multirow{2}{*}{\textbf{Key Contributions}}} & \multicolumn{1}{c|}{\multirow{2}{*}{\textbf{Limitations}}}  \\ 
		
		\cline{2-3}
		{} & \textbf{TinyML} & \textbf{LargeML} & {} & \\
		\hline
		\hline

        \multirow{3}{*}{\cite{dutta2021tinyml}} 
		& \multirow{3}{*}{\checkmark}  &  
		& Highlights hardware-software co-design for optimizing TinyML frameworks on resource-limited devices and reviews ML algorithms optimized for low-power deployment.
		& Focuses on TinyML and 5G, without discussion of integration with LargeML in 6G systems.
		\\ \hline
		
		\multirow{3}{*}{\cite{abadade2023tinyml}} 
		& \multirow{3}{*}{\checkmark} & 
		& Introduces a classification system for TinyML applications based on an extensive literature review, exploring benefits and use cases.
		& Does not address TinyML--LargeML integration or the specific challenges of 6G integration.
		\\ \hline
		
		\multirow{2}{*}{\cite{ray2022review}} 
		& \multirow{2}{*}{\checkmark} & 
		& Surveys TinyML frameworks, development tools, enablers, and applications, and outlines current research challenges.
		& Lacks discussion of TinyML--LargeML integration in the context of 6G networks.
		\\ \hline
		
		\multirow{2}{*}{\cite{rajapakse2023intelligence}} 
		&  \multirow{2}{*}{\checkmark} & 
		& Proposes a taxonomy for reformable TinyML solutions and evaluates the suitability of TinyML layers for reformability.
		& Does not explore TinyML--LargeML integration in the context of 6G networks.
		\\ \hline
		
		\multirow{3}{*}{\cite{tsoukas2024review}} 
		& \multirow{3}{*}{\checkmark} & 
		& Reviews neural network (NN) compression techniques and categorizes TinyML hardware and software resources.
		& Examines the foundational aspects of TinyML, with limited discussion on integration with LargeML or implications for 6G environments.
		\\ \hline

        \multirow{2}{*}{\cite{capogrosso2024machine}} 
		& \multirow{2}{*}{\checkmark} & 
		& Reviews TinyML research, workflows, algorithms, and development tools, with an emphasis on recent ML advances. 
		& Lacks analysis of integration with LargeML in the context of 6G networks.
		\\ \hline

        \multirow{3}{*}{\cite{raiaan2024review}} 
		&  &  \multirow{3}{*}{\checkmark}
		& Reviews transformer architectures in LLMs, discussing training methods, applications, societal impact, and deployment challenges.
		& Does not address 6G-related challenges or TinyML--LargeML integration for enhancing 6G capabilities.
		\\ \hline 

		\multirow{3}{*}{\cite{chang2024survey}} 
		&  &  \multirow{3}{*}{\checkmark}
		& Examines LLM evaluation methods, benchmarks, and performance across domains.
		& Omits discussion of ML integration and analysis of challenges in 6G applications. 
		\\ \hline

		\multirow{3}{*}{\cite{han2024parameter}} 
		&  &  \multirow{3}{*}{\checkmark}  
		& Analyzes fine-tuning algorithms for LargeML, detailing performance metrics and deployment scenarios.
		& Lacks the investigation of fine-tuning algorithms with TinyML for 6G applications.
		\\ \hline

        \multirow{3}{*}{\cite{zhao2024explainability}} 
		&  &  \multirow{3}{*}{\checkmark}
		& Proposes a taxonomy of explainability techniques for LLMs, with evaluation metrics and implementation challenges.
		& Concentrates on LLM explainability without addressing its integration with TinyML in collaborative 6G systems.
		\\ \hline

        \multirow{2}{*}{\cite{min2023recent}} 
		&  &  \multirow{3}{*}{\checkmark}
		& Outlines architectures, techniques, and challenges in PLMs.
		& Focuses on NLP applications of PLMs; lacks discussion of TinyML integration in 6G systems.
		\\ \hline\hline

		\multirow{8}{0.9cm}{This work} 
		& \multirow{8}{*}{\checkmark} & \multirow{8}{*}{\checkmark}  
		& Offers the first survey of the integration of TinyML and LargeML for 6G systems. Specifically, (\textit{i}) Introduces fundamentals of TinyML and LargeML;
        (\textit{ii}) Discusses motivations and requirements for integrating TinyML and LargeML in 6G;
        (\textit{iii}) Proposes bidirectional integration frameworks for enhancing future network performance and capabilities;
        (\textit{iv}) Explores potential applications of integrated TinyML--LargeML systems;
        and (\textit{v}) Identifies challenges and future research directions for TinyML--LargeML integration. 		
		 & {}
		\\ \hline
  
        \hline
        
	\end{tabular}
    }
\end{table*}
 
The previous discussion reveals a significant gap in surveys addressing the integration of TinyML and LargeML in the context of 6G and beyond. Although recent advancements in LargeML \cite{zhuang2023foundation, ren2024advances} are vital for developing efficient 6G networks, they have not been thoroughly analyzed. Most studies focus on either TinyML or LargeML \cite{wang2020survey}, lacking a cohesive examination of their combined potential to tackle emerging network challenges. This survey aims to fill that gap by exploring the integration of TinyML and LargeML for 6G and future networks. A key contribution of this work is a thorough exploration of the motivations and requirements for such integration. Our contributions are summarized below:
\begin{enumerate}
    \item \textbf{Background}: This work first provides the background of TinyML and LargeML, highlighting their individual advances and potential for cooperation.
    \item \textbf{Integration requirements}: We outline the motivations and requirements for integrating TinyML and LargeML to emphasize their importance for 6G networks. 
    \item \textbf{Efficient solutions}: We propose and discuss efficient bidirectional integration solutions aimed at enhancing the performance and capabilities of future networks.
    \item \textbf{Applications of integrated systems}: We conduct an extensive review of the applications of integrated TinyML and LargeML systems across various domains to demonstrate their practical benefits and potential impacts.
    \item \textbf{Discussion of challenges and future research}: From our extensive review, we identify critical challenges and propose several potential directions for future research, including resource allocation, communication efficiency, interplay with LargeML, and standardization.
\end{enumerate}



\section{Overview of TinyML and LargeML}
\label{Sec:Overview} 

\subsection{Overview of TinyML}
TinyML is a lightweight, resource-constrained class of ML designed for environments with limited memory and processing power \cite{Lin2023Oct}. TinyML supports diverse communication tasks, and its characteristics are summarized in Table~\ref{tab:TinyML}.  
\begin{table*}[!t]
    \centering
    \caption{TinyML classifications, hardware devices, environmental deployment, and software tools.}
    \label{tab:TinyML}
    \begin{tabular}{p{0.269\textwidth}|p{0.215\textwidth}|p{0.225\textwidth}|p{0.2\textwidth}}
    \hline 
       \multicolumn{1}{c|}{\textbf{Classification}} & \multicolumn{1}{c|}{\textbf{Hardware Device} \cite{capogrosso2024machine}} & \multicolumn{1}{c|}{\textbf{Environmental Deployment} \cite{abadade2023tinyml} } & \multicolumn{1}{c}{\textbf{Software Tools} \cite{rajapakse2023intelligence}}    \\
       \hline \hline 
       \begin{minipage}[t]{0.269\textwidth} 
          \begin{itemize}[leftmargin=2pt, noitemsep, topsep=0pt]
          \item \textbf{Model size}: (1) Compression methods (pruning, quantization, knowledge distillation), (2) Memory/storage design (kilobytes to megabytes), (3) Working modes (centralized, distributed, decentralized).
          
          \item \textbf{Architecture}: (1) Convolutional neural network (CNN) for image processing, (2) Recurrent neural network (RNN) for sequential data, (3) Fully connected networks for certain tasks, (4) Transformer \cite{vaswani2017attention} for data generation, context awareness, translation, and summarization. 
          
          \item \textbf{Training data}: (1) Quality assurance and data labeling, (2) Data augmentation (rotation, scaling, flipping), and (3) Fine-tuning pre-trained models on large datasets.

          \item \textbf{Algorithms}: {Supervised, weakly supervised, unsupervised, self-supervised, meta-learning, continual learning, deep reinforcement learning (DRL), transfer learning (TL), split learning (SL), and federated learning (FL) \cite{abadade2023tinyml} (readers may refer to \cite{mohri2018foundations,bishop2023deep,wani2025advances,wang2021introduction,yang2019federated} for additional foundations and core concepts).}
         \end{itemize} 
        \end{minipage}
        
       &
       \begin{minipage}[t]{0.215\textwidth} 
          \begin{itemize}[leftmargin=4pt, noitemsep, topsep=0pt]
          \item \textbf{Central processing unit (CPU)}: Arm Cortex M family offers ultra-low-power operation but limits to parallel processing.
          
          \item \textbf{Graphics processing unit (GPU)}: Supports parallel computing and optimized sequential processing of large datasets (e.g., NVIDIA Jetson family, AMD, Intel, Arm), but is not cost-effective.
          
          \item \textbf{Field-programmable gate array (FPGA)}: Enables model customization and hardware-level acceleration for complex operations and large dataset access, but lacks broad library support.
          
          \item \textbf{Tensor processing unit (TPU)}: Accelerates ML workloads, particularly those involving NNs, with two representations of Edge TPUs and Cloud TPUs. 
         \end{itemize} 
        \end{minipage}
    
       &
       \begin{minipage}[t]{0.225\textwidth} 
      \begin{itemize}[leftmargin=6pt, noitemsep, topsep=0pt]
      \item \textbf{Atmospheric}: Low-power sensors to monitor air quality in urban, forest, and industrial areas, detecting pollutants by measuring particulate matter, carbon dioxide levels, and others.
      
      \item \textbf{Hydrosphere}: Sensors in aquatic environments to detect pollutants, measure pH, and monitor temperature changes.

      \item \textbf{Biosphere}: Device used for wildlife conservation, monitoring biological activity, or classifying sounds (e.g., animal habitats and anti-poaching surveillance).
      
      \item \textbf{Lithosphere}: Applied in agriculture for environmental monitoring (e.g., soil moisture, nutrient levels, pests) or earthquake detection. 
      \item \textbf{Human-related}: Wearable health monitor (e.g., blood pressure) and assistive technologies for humans (e.g., mobility). 
     \end{itemize} 
    \end{minipage}
       &  
       \begin{minipage}[t]{0.21\textwidth} 
      \begin{itemize}[leftmargin=8pt, noitemsep, topsep=0pt]
      \item \textbf{Mobile device}: TensorFlow Lite for model optimization and PyTorch for programming.
      
       \item \textbf{Embedded devices}: TensorFlow Lite for model optimization, PyTorch for programming, and Qeexo AutoML for end-to-end customization.
            
       \item \textbf{Edge devices}: PyTorch for programming, NanoEdgeAIStudio for low/no-code solution, and Imagimob/Edge Impulse for end-to-end customization.
       
       \item \textbf{Microcontroller}: Arm NN for network architecture optimization, MinUn for bandwidth/memory management, STM32CubeA and uTensor for pre-trained model conversion, and NXP eIQ as a development environment.
     \end{itemize} 
    \end{minipage}
       \\
       \hline 
    \end{tabular}
\end{table*}

\begin{figure}[!t]
\centering
\includegraphics[width=.95\linewidth]{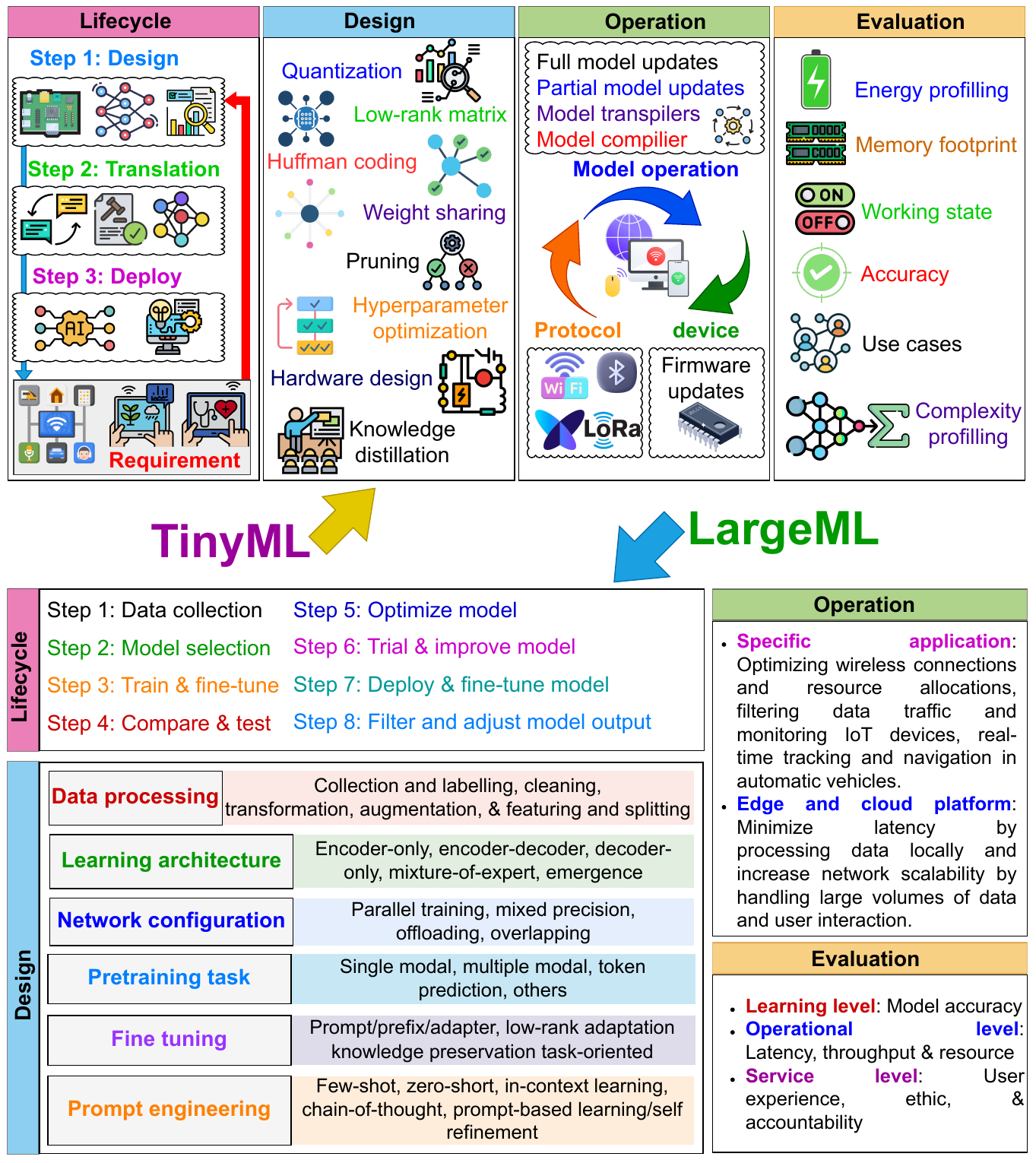}
\caption{Overview of TinyML and LargeML, including their life cycle, design, operation, and performance evaluation.}
\label{Fig:TinyML-LargeML}
\end{figure}

As \cite{wu2025consolidating}, the TinyML life cycle has three stages:
(\textit{i}) \textit{\textbf{Stage 1}}: Data collection from peripheral devices (sensors or production environments);
(\textit{ii}) \textit{\textbf{Stage 2}}: Model deployment using quantization and compression to address hardware heterogeneity and resource constraints;
and
(\textit{iii}) \textit{\textbf{Stage 3}}: Model operationalization, converting trained models into interpretable forms for target devices.
TinyML relies on application-specific data formats, prioritizing model customization for device capacity and efficient data transmission (see Fig.~\ref{Fig:TinyML-LargeML} for a short overview).

\subsubsection{TinyML Design} \label{TinyML_Design}
While no standard design guidelines exist, several best practices are applied to guide development.

\textbf{Quantization}: Reduces the memory footprint and computation by representing parameters in lower precision (e.g., 8-bit integers versus 32-bit floats) \cite{rokh2023comprehensive}. It can be applied during training (quantization-aware training) or after training (post-training quantization). Variants include uniform and non-uniform quantization for optimizing accuracy, symmetric or asymmetric schemes to reduce quantization error, and Bayesian inference to minimize bit encoding requirements.

\textbf{Pruning}: Removes redundant weights, neurons, or layers to shrink NN size \cite{zhu2017prune}. Pruned models typically require fine-tuning to restore accuracy. Common strategies include threshold-based pruning, in which weights below a specified magnitude are discarded. From differences in the model weight matrix, pruning is classified as unstructured (removal of individual weights) or structured (removal of neurons, filters, rows, or columns), guided by redundancy or regularization metrics.

\textbf{Low-rank matrix decomposition}: Approximates large weight matrices via techniques like singular value decomposition, with lower-rank components. The resulting factorized forms contain fewer parameters, reduce computational complexity, and simplify matrix operations \cite{udell2016generalized}. Albeit this technique can be computationally intensive and more difficult to implement, as the decomposed network requires additional fine-tuning to recover accuracy loss, similar to model pruning.

\textbf{Weight sharing}: Reduces parameters across model components to cut storage and computation \cite{ott2020learning}. For example, shared weights can be arranged by row or column within a weight matrix. However, designing effective sharing configurations is difficult and can lead to unpredictable performance.

\textbf{Algorithm-architecture co-design}: Jointly optimizes algorithms with specialized hardware (e.g., TPUs, Cortex-M7, Cortex-M4, Cortex-M0+, and eDMPv1), for inference efficiency \cite{le2023efficient}. This joint optimization also enhances computational performance and energy efficiency.

\textbf{Huffman coding}: Compresses model parameters using variable-length encoding based on symbol frequency \cite{moffat2019huffman}. When applied to model weights and parameters, this technique can reduce overall model size through efficient encoding.
   
\textbf{Hyperparameter optimization}: 
Tunes model parameters, such as learning rate, batch size, activation functions, number of neurons, and network depth, typically via grid search, random search, Bayesian optimization, Hyperband (which dynamically allocates resources to promising configurations while terminating underperforming configurations early), genetic algorithms, tree-structured Parzen estimators, and hyperdimensional computing \cite{feurer2019hyperparameter}. These techniques aim to improve model performance by minimizing memory and computational overhead while ensuring user-defined accuracy.
    
\textbf{Knowledge distillation}: Transfers knowledge from a larger “teacher” model to a smaller “student” via (\textit{i}) response-based, where students mimic the teacher's final output; (\textit{ii}) feature-based, where students replicate intermediate feature representations; and (\textit{iii}) relation-based, where students learn the structural relationships within the teacher model \cite{yang2023categories}. 

\subsubsection{TinyML Operation}
From an operational standpoint, TinyML can be categorized into three main levels: firmware updates, model operation, and communication protocols \cite{huang2024riot}. 

\textbf{Firmware updates}: Replace the entire software stack, including the TinyML model and system components. While comprehensive, this process is resource- and time-intensive. 
Over-the-air (OTA) delivery via lightweight M2M protocols (push/pull) is common  \cite{sudharsan2022ota}. Tools such as RIOT-ML facilitate secure OTA updates, enabling remote, unattended deployment.

\textbf{Model operations}: Several strategies optimize TinyML deployment \cite{huang2024riot}. \textit{Full model updates} replace the entire ML model via OTA delivery, ideal for major changes, ensuring devices run the latest version. \textit{Partial model updates} modify only specific layers or parameters, reducing network traffic and power use, suitable for fine-tuning on constrained devices \cite{ostrovan2022tinyml}. \textit{Model transpilers}  (e.g., TensorFlow $\to$ TensorFlow Lite Micro), improve cross-hardware compatibility and resource efficiency \cite{david2021tensorflow}. \textit{Model compilers} (e.g., uTensor, TinyTS) generate optimized binaries using quantization and pruning, reducing size and improving inference efficiency, fitting within tight memory and computational limits.

\textbf{Communication protocols}: TinyML relies on specialized protocols to optimize data transfer under hardware constraints in IoT ecosystems \cite{capogrosso2024machine}. 
For \textit{intra-device} links, inter-integrated circuit and serial peripheral interface protocols support efficient data exchange between MCU-peripheral exchange. 
For \textit{device-cloud} links, message queuing telemetry transport offers lightweight transfer in low-bandwidth, high-latency environments. 
\textit{Short-range} links are met by low-power wide-area network (LoWPAN) protocols (Bluetooth, Zigbee, and 6LoWPAN) for wearable and smart sensors, while Wi-Fi supports higher data rates. 
For \textit{long-range, low-power} links, long-range wide area network excels in remote applications like agriculture and smart cities, offering energy-efficient, kilometer-scale coverage. 
Sigfox, designed for ultra-narrowband IoT and M2M tasks, transmits small, infrequent data packets on unlicensed bands, ensuring low-cost, secure deployments. 
Similarly, Narrowband IoT (3GPP) supports low-data-rate, energy-efficient communication in licensed/unlicensed spectra, enabling scalable, reliable connectivity for massive IoT deployments in challenging environments like underground or indoor settings.

\subsubsection{Performance Evaluation}
Given the diversity of application contexts and strict resource limitations, performance evaluation of TinyML models requires the consideration of both device constraints and learning model characteristics.   

\textbf{Memory Footprint}: TFLite Micro targets microcontrollers with limited ROM and RAM, using quantization and pruning to reduce model size with minimal accuracy loss \cite{cai2020tinytl}. Evaluating ROM and RAM consumption is crucial for assessing TinyML deployments on IoT devices.

\textbf{On/Off characteristics}: TinyML components (sensors, protocols, inference engines) switch between sleep and active power states to save energy \cite{david2021tensorflow}. Optimizing the duration and frequency of these transitions minimizes active time, enhancing energy efficiency while preserving functionality.

\textbf{Complexity profiling}: This refers to the analysis of the computational demands of TinyML models. Floating-point operations are more computationally intensive than quantized int8 operations \cite{capogrosso2024machine}, which reduce weight and activation precision to shrink model size and accelerate inference. Complexity profiling helps quantify the trade-offs between model size, accuracy, and inference performance \cite{Lin2023Oct}.

\textbf{Energy profiling}: Energy profiling measures CPU cycles, latency, and total energy consumption during inference \cite{sabovic2023towards}. Floating-point models typically consume more energy than quantized int8 models due to the greater computational overhead. Thus, it helps identify energy-efficient TinyML models.

\textbf{Model Evaluation Metrics}: Several metrics are used to evaluate TinyML model performance: Accuracy, mean absolute error (MAE), root mean square error (RMSE), R-Squared, precision, recall, F1-score, and cross-entropy loss.

\textbf{Deployment Robustness}: 
Deployment durability ensures the reliability of the TinyML model under a variety of environmental conditions. Two evaluation methods are used \cite{huang2024riot}: \textit{per-model evaluation} assesses performance in controlled pre-deployment tests, while \textit{per-operator evaluation} monitors adaptability in real-world environments. Durability testing encompasses diverse datasets, hardware, and environments to ensure consistent performance.

\subsection{Overview of LargeML}
LargeML is increasingly integrated into wireless networks to enable autonomous operations via cognitive modules for resource allocation, channel estimation, and mobility management \cite{xu2024large}.
Its life cycle consists of eight stages:
(\textit{i}) \textbf{\textit{Stage~1}}: Data collection and pre-processing from multiple domains for high-quality training corpus;
(\textit{ii}) \textbf{\textit{Stage 2}}: Model architecture design regarding layer types, parameters, and training algorithms;
(\textit{iii}) \textbf{\textit{Stage 3}}: Training through pre-training on broad data and fine-tuning for task-specific applications;
(\textit{iv}) \textbf{\textit{Stage 4}}: Evaluation with benchmarks, refining architecture, hyperparameters, and datasets;
(\textit{v}) \textbf{\textit{Stage 5}}: Compression to reduce memory and latency while retaining performance;
(\textit{vi}) \textbf{\textit{Stage 6}}: Testing and optimization under controlled conditions for inference efficiency;
(\textit{vii}) \textbf{\textit{Stage 7}}: Deployment with continuous monitoring, fine-tuning, and adaptation;
and (\textit{viii}) \textbf{\textit{Stage~8}}: Risk mitigation using adversarial training, fairness constraints, and curated datasets, addressing bias, privacy, security, and misuse concerns.

\subsubsection{LargeML Design} The design process for LargeML covers six key areas, outlined below. 

\paragraph{Data Processing} 
Training LargeML models depends on diverse IoT data, as quality impacts prediction accuracy. The process involves three key steps.

\textbf{Collection and labeling}: Data is gathered from heterogeneous sources (real-time sensors, logs, user feedback) and annotated with relevant features (e.g., signal strength, noise, mobility patterns \cite{le2024applications}). Accurate labeling prevents bias and ensures models generalize across operational scenarios \cite{huang2024data}.
        
\textbf{Cleaning, transformation, and augmentation}: Raw data is cleaned by removing duplicates, correcting errors, and handling missing values. It is then transformed through normalization, encoding, and feature engineering, and represented in formats suitable for models, such as token-based input for text, tree structures for hierarchical relationships, and pixel-based representations for visual data \cite{chang2024survey}. Augmentation methods, like rotation, flipping, and noise injection, enhance sample diversity and robustness, resulting in a clean and well-structured dataset for downstream tasks.
    
\textbf{Featuring and splitting}: 
It involves feature selection to identify high-impact variables, feature extraction to derive new indicators (e.g., signal variation rates) and capture deeper patterns, as well as data partitioning for training, validation, and test sets to ensure robustness and generalizability \cite{muraina2022ideal}. 

\paragraph{Learning Architecture} 
Future 6G networks aim to support integrated sensing and communication (ISAC), task-oriented communications, and digital twins, requiring LargeML models for efficient real-time data processing  \cite{xu2024large}. Selecting suitable learning architectures is challenging due to application-specific constraints. Transformers, leveraging attention mechanisms, outperform RNNs in sequence data processing \cite{vaswani2017attention}. They use an encoder-decoder structure with stacked sub-layers: encoders create context-aware representations, positional embeddings preserve token order, and decoders apply masked or multi-head self-attention to assess token relevance. Aside from three typical architectures (\textbf{Encoder only} \cite{lee2018pre}, \textbf{Encoder-Decoder} \cite{vaswani2017attention}, and \textbf{Decoder only} with types of causal decoders \cite{radford2019language} and decoders \cite{zhang2022examining}), below are new variants of transformer architectures.

\textbf{Mixture of experts (MoE)}: MoE activates only subsets of model weights, improving efficiency as in Switch Transformer \cite{fedus2022switch} and GLaM \cite{du2022glam}.
Performance scales with more experts or larger parameter size, but unstable training can occur due to complex routing operations. This issue can be mitigated by precision tensors, initialization adjustments, and hybrid resource-conditional computation \cite{tran2025revisiting}.

\textbf{Emergence architectures}:
Standard transformers face quadratic complexity with sequence length.
Motivated by this, parameterized state-space models \cite{gu2021efficiently} integrate RNN- and CNN-like properties: they process outputs recursively (like RNNs) while encoding sequences in parallel (like CNNs).
Techniques such as fast Fourier transform and chunk-wise recurrence can further improve efficiency.

\paragraph{Network Configuration} 
Network setup strongly affects training performance.
Common strategies address parallelism, precision, memory, and computation \cite{liu2024understanding}.

\textbf{Parallel training}: Data parallelism synchronizes GPUs using a parameter server/all-reduce gradients; pipeline parallelism splits models across devices; and model parallelism distributes layers/neurons, exchanging intermediate results.

\textbf{Mixed precision and offloading}: Mixed precision stores gradients in single-precision (32-bit) while training in half-precision (16-bit) to prevent underflow and ineffective parameter updates. Offloading optimizer states from the GPU to multiple CPUs alleviates memory bottlenecks.

\textbf{Overlapping and check-pointing}: Overlapping fetches parameters while computing on previously loaded ones, and checkpointing stores only minimal activations and recomputes the rest to trade memory for compute.

\paragraph{Pre-training Tasks} 
Pre-training equips models with general knowledge that minimizes the need for task-specific training. It involves large-scale data collection, self-supervised learning to capture foundational representations, and fine-tuning for target tasks. LargeML models are pre-trained on extensive corpora to learn universal syntactic and semantic features, which are later adapted for downstream applications. Common pre-training workflows are outlined below \cite{wei2024overview}.

\textbf{Single-modal}: Trains on a single modality (e.g., text, images, audio) to capture domain-specific patterns. 
For instance, pre-training on large text corpora enables strong performance in sentiment analysis, speech classification, and translation.

\textbf{Multi-modal}: Combines different modalities to learn richer representations.
For example, training image captions with visuals improves tasks such as captioning, visual question answering, and cross-modal retrieval.

\textbf{Token prediction}: Trains models to predict missing or next tokens, enhancing contextual understanding. Masked language modeling (e.g., BERT) predicts hidden random tokens, while auto-regressive modeling (e.g., GPT) generates tokens sequentially for coherent outputs.

\paragraph{Fine-tuning} 
After pre-training, a model is fine-tuned on a smaller, task-specific dataset to adjust its parameters for the target application. This process builds on the knowledge acquired during pre-training. In addition to the TinyML strategies discussed in Section~\ref{TinyML_Design}, this section reviews the fine-tuning techniques relevant to LargeML systems \cite{hou2024large}.

\textbf{Prompt tuning}: Prompt tuning introduces trainable tokens (prompts) to guide the model's output without modifying its architecture or parameters.

\textbf{Prefix tuning}: This technique prepends a set of trainable tokens (prefixes) to the model's input and intermediate layers. These prefixes steer the model's attention and interpretative processes without updating the core parameters, enabling task adaptation with low computational overhead.

\textbf{Adapter tuning}: Adapter tuning inserts small NN modules (adapters) into the model's architecture. During fine-tuning, only the adapters are trained to capture task-specific information \cite{houlsby2019parameter}, while the original model remains unchanged.

\textbf{Low-rank adaptation (LoRA):} LoRA enhances transformer models by introducing trainable low-rank matrices to selected layers, keeping original weights frozen \cite{hu2022lora}. Only lightweight matrices are trained to approximate weight updates, thereby reducing memory and computational costs.

\textbf{Knowledge preservation}: 
This technique adapts a pre‑trained model to new tasks while preserving its valuable pre-existing knowledge, mitigating \textit{catastrophic forgetting} through causal‑aware methods and PEFT techniques such as LoRA and Half Fine‑Tuning. The result is a model that stays effective across tasks without overfitting to new data.

\textbf{Task orientation}: This technique adapts pre-trained models to specific tasks (e.g., dialogue systems) by fine-tuning the model's parameters on task-specific data. The incorporation of techniques such as adapter layers and prefix tuning reduces the overall parameter count for training, improving efficiency and lowering resource demands.

\paragraph{Prompt Engineering}
After fine-tuning, prompt engineering refines how the model interprets and responds to inputs according to task requirements and application contexts. Common types include multi-turn instructions, multiple modeling, task planning, tool augmented alignment, and retrieval augmentation. The most widely used prompting techniques in wireless networks are described below \cite{xu2024large}.

\textbf{Few-shot prompting}: Enables the model to generalize from a small number of examples. It is particularly useful when large annotated datasets are unavailable, allowing the model to perform new tasks with minimal additional data.

\textbf{Zero-shot prompting}: Exploits the model's pre-trained knowledge to perform tasks without prior examples or specific training. The model generates outputs based solely on instructions, drawing from its existing knowledge. In contrast to few-shot prompting, zero-shot prompting provides no examples to guide the model's output.

\textbf{In-context learning}: 
Enables models to perform tasks from prompts alone, without parameter updates. By analyzing examples in the prompt, the model infers task structure and applies it to new inputs, e.g., detecting network intrusions, adapting to threats, or managing resources in real time.

\textbf{Chain-of-thought}: Enhances reasoning by breaking tasks into intermediate steps, ensuring logical and coherent decision-making. Useful in applications where stepwise analysis accelerates repairing and improves resilience, e.g., fault diagnosis.

\textbf{Prompt-based learning}: Leverages pre-trained knowledge to generate accurate, contextually relevant responses without further training or parameter adjustments. Well-designed prompts guide task execution, e.g., predicting maintenance needs and preventing outages by identifying failure patterns.

\subsubsection{LargeML Operation} LargeML follows a multi-stage process: large-scale data collection and pre-processing, initial training on high-performance hardware, fine-tuning for specific tasks, and post-deployment adaptation through real-time inputs and feedback. This ensures both accuracy and long-term relevance.
Its operation varies across domains and input data characteristics from sensors and devices \cite{xu2024large}. 
In wireless communication, LargeML optimizes connectivity and manages network traffic across mobile, Wi-Fi, and satellite links. 
In the IoT domain, it processes data from smart homes, connected vehicles, and healthcare monitors to enable intelligent automation and real-time decision-making. 
For example, it serves as an AI control hub in smart homes, supports navigation and diagnostics in vehicles, and enables remote monitoring and personalized care in healthcare.
Deployment strategies include edge-based processing to reduce latency and cloud integration for scalability \cite{hou2024large}.
Continuous monitoring and updates based on user feedback and new data sustain model performance across applications. Integrated into wireless and IoT systems, LargeML supports smart city infrastructure (e.g., traffic and energy optimization), predictive industrial maintenance, and smart agriculture through environmental monitoring.

\subsubsection{Performance Evaluation}
The performance of LargeML can be assessed at three levels, described below.

At the \textbf{learning level}, the focus is on accuracy, output quality, and efficiency. Accuracy is assessed using precision, recall, F1 score, and area under the receiver operating characteristic (AUC), which measure classification and discrimination ability \cite{neubius}. Output quality is gaged using perplexity (confidence in predictions), bilingual evaluation understudy (BLEU), and recall-oriented understudy for gisting evaluation (ROUGE), which compare generated text against reference outputs.


At the \textbf{operational level}, performance is measured by typical metrics, such as latency (response time), throughput (requests processed per unit time), and resource use (computational cost), ensuring scalable and reliable operation. 

At the \textbf{service level}, evaluation focuses on user experience, ethics, and accountability \cite{chang2024survey,zhao2024explainability}. 
User experience includes user satisfaction (perception feedback), engagement (interaction depth), and retention rate (continued usage's proportion).
Ethical and accountability metrics assess fairness (bias reduction), transparency (explainable decisions), and ethical compliance (privacy, security, and accountability standards).

\section{Motivations and Requirements for TinyML-- LargeML Integration in 6G Networks}
\label{Sec:Motivation}

\subsection{Challenges in TinyML}\label{Sect:IIIA}


\subsubsection{Resource Constraints and Energy Efficiency}
TinyML devices typically operate with limited processing power, memory, and energy resources, complicating the deployment of complex models. For instance, microcontrollers often lack DRAM and an operating system, with memory capacities ranging from a few hundred kilobytes to several megabytes \cite{lin2022device,abadade2023tinyml}. Such limitations preclude the use of LargeML without aggressive compression, often forcing trade-offs between accuracy and feasibility. Furthermore, many TinyML devices are battery-powered, requiring energy-efficient models that minimize inference overhead to extend device lifespan. Achieving low-latency inference while maintaining energy efficiency remains a key challenge, especially for mission-critical 6G applications such as autonomous driving and industrial automation \cite{tuama2016camera}.

\subsubsection{Model Deployment and Data Pre-processing}
The deployment of ML models on TinyML devices necessitates compression techniques such as pruning, quantization, and knowledge distillation \cite{TinyML_market}. Although these approaches reduce memory and computation costs, they may compromise accuracy, particularly when parameters essential for prediction are removed. Quantized optimization introduces additional difficulties due to mixed-precision tensors and the absence of batch normalization layers \cite{ioffe2015batch}. Beyond model compression, pre-processing tasks such as normalization, augmentation, and feature extraction require computational and memory resources that are typically unavailable on TinyML devices. This makes handling complex datasets, such as high-resolution images, particularly challenging.

\subsubsection{Reliability, Privacy, and Security}
TinyML devices suffer from limited numerical precision and fabrication‑related errors, which can undermine system reliability, especially in mission‑critical settings like healthcare, where failures may endanger patient safety \cite{tsoukas2021review}. They also face significant privacy and security risks: although they often handle sensitive audio or visual data, their constrained compute capacity prevents the use of stronger protections, such as differential privacy or secure multiparty computation, leaving them more exposed to data leakage and adversarial attacks \cite{shabir2023toward}.

\subsection{Challenges in LargeML}\label{Sect:IIIB}

\subsubsection{Resource and Infrastructure Demands}
Training and inference of LargeML require specialized hardware such as GPUs or TPUs, along with extensive RAM and storage for managing parameters and activation states \cite{koubaa2023gpt}. Training often spans days or weeks, incurring high financial and energy costs. Such requirements limit accessibility, particularly for small and medium enterprises or researchers in developing regions, where infrastructure and expertise may be insufficient \cite{grace2024thousands}.

\subsubsection{Latency and Scalability Issues}
Due to their size and complexity, LargeML models are prone to high inference latency. Offloading computation to cloud servers introduces further network delays, which are unacceptable for latency-sensitive 6G applications such as augmented reality, virtual reality, and autonomous vehicles \cite{mohammed2019artificial}. Scaling LargeML is further hindered by the need to manage vast, heterogeneous, and often non-independent and identically distributed (non-IID) datasets across distributed infrastructures. This increases the risks of bottlenecks, degraded model performance, and challenges in real-time processing \cite{barmer2021scalable}.

\subsubsection{Deployment, Interpretability, and Adaptability}
Deploying LargeML involves costly and complex system integration, including hardware configuration, software optimization, and application-specific tuning. Moreover, these models often operate as “closed boxes,” with opaque decision-making processes. This lack of interpretability undermines trust in mission-critical domains such as healthcare and finance, where explainability and accountability are essential \cite{gdpr2016general}. Effective fine-tuning is also challenging: while adaptation to new data domains is necessary, it requires careful balancing to avoid overfitting or underfitting \cite{ding2023parameter}.

\subsubsection{Privacy and Security Concerns}
LargeML models typically rely on centralized data aggregation, which concentrates sensitive information in repositories vulnerable to cyberattacks \cite{dilmaghani2019privacy}. Inference tasks may also expose confidential data, complicating real-time privacy protection. Furthermore, these models are susceptible to threats such as adversarial attacks, model extraction, and reverse engineering, which can compromise both model integrity and the confidentiality of training data.

\subsection{The Vision for 6G} \label{Sect:IIIC}

\begin{figure}[!t]
\centering
\includegraphics[width=.8\linewidth]{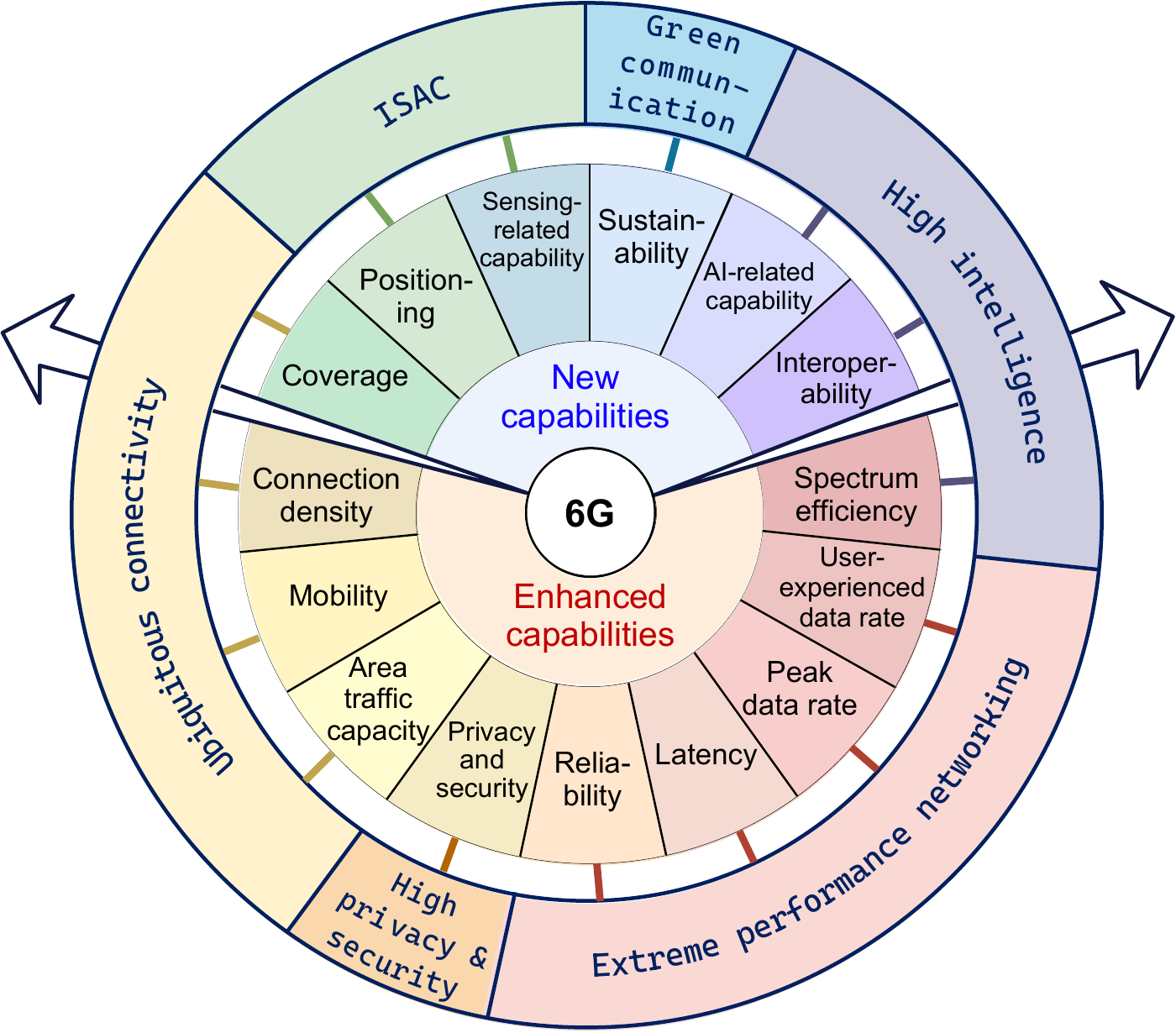}
\caption{Envisaged 6G capabilities and pivotal requirements.}
\label{Fig:6G_requirements}
\end{figure}

IMT-2030 framework identifies six overarching requirements for 6G networks: ubiquitous connectivity, extreme performance, high intelligence, strong security, green communication, and efficient ISAC (see Fig.~\ref{Fig:6G_requirements}) \cite{liu2023beginning}. Standalone approaches (whether TinyML or LargeML) are insufficient to satisfy these requirements, necessitating a bidirectional integration paradigm. TinyML supports local data processing at the edge, while LargeML leverages high-capacity infrastructure for global-scale optimization. Together, they enable complementary capabilities aligned with the 6G vision.

\subsubsection{Ubiquitous Connectivity}
6G must support ultra-dense networks, with connection densities up to $10^6-10^8$~devices/km$^2$ and data volumes five times higher than 5G \cite{liu2023vision}. TinyML enables edge devices to process local data in real time, reducing network load, while LargeML, deployed across a \textit{ubiquitous server set} (cloud, edge, mobile devices) \cite{lin2024split}, aggregates distributed information to optimize resource allocation and mobility management. This integration supports global coverage across terrestrial, aerial, satellite, and underwater systems, with intelligent handover strategies for high-mobility platforms \cite{dao2021survey,dao2023neglected}.

\subsubsection{Extreme Performance Networking}
Target metrics for 6G include latencies as low as 0.1 ms and reliability levels up to 99.99999, with peak data rates projected to reach 1 Tbps \cite{ITU_R_M2160}. Bidirectional integration enhances performance by reducing latency at the edge (TinyML) while maintaining complex server-side analytics (LargeML). Model splitting, model transfer, hierarchical distribution, and lightweight fine-tuning enable real-time collaboration across devices and servers, improving throughput, resilience, and accuracy.

\subsubsection{High Intelligence}
6G envisions distributed intelligence across networks, combining localized processing and global decision-making \cite{ITU_R_M2160}. TinyML provides real-time inference for edge devices, while LargeML ensures resource-intensive optimization and personalization. This collaboration supports bandwidth-heavy applications such as ISAC and holographic communication \cite{liu2023vision}, where AI-driven spectrum allocation and cognitive technologies improve efficiency and responsiveness.

\subsubsection{Strong Privacy and Security}
As 6G interconnects billions of devices, ensuring data confidentiality and system robustness is paramount. TinyML enhances privacy by performing local pre-processing and transmitting only essential features, thereby reducing exposure of raw data. Concurrently, LargeML systems, supported by high-performance servers, can conduct sophisticated threat detection and adaptive defense using aggregated data from multiple devices \cite{huawei20226g}.

\subsubsection{Green Communications}
With the proliferation of connected devices, energy efficiency is central to sustainable 6G systems. TinyML reduces reliance on energy-intensive servers by enabling low-power, on-device processing. LargeML complements this by optimizing network-wide resource management, minimizing redundant processing, and reducing overall CO$_2$ emissions \cite{huang2019survey}. Bidirectional integration also supports predictive maintenance of network infrastructure and the use of energy-harvesting technologies.

\subsubsection{Efficient ISAC}
ISAC represents a defining requirement of 6G, merging communication, sensing, and localization into a unified framework \cite{liu2022integrated}. TinyML supports on-device feature extraction and preliminary sensing tasks, while LargeML performs advanced analysis for object detection, imaging, and mapping. This integration enhances precision, reduces false positives, and enables real-time navigation and environmental awareness, being key for autonomous systems and smart cities.


\subsection{Discussions and Lessons Learned} 
The transition from 5G to 6G creates significant opportunity to integrate TinyML and LargeML to enhance AI capabilities in next-generation networks.
Section~\ref{Sect:IIIA} discusses the key challenges of TinyML, while Section~\ref{Sect:IIIB} examines the limitations of LargeML.
As highlighted in Section~\ref{Sect:IIIC}, a hybrid framework that combines TinyML and LargeML leverages the strengths of distributed and centralized learning, forming a promising paradigm aligned with the 6G vision.
In essence, TinyML enables edge-level data preprocessing to reduce bandwidth consumption and latency, whereas LargeML performs deeper analysis for high-accuracy insights. 
This bidirectional integration, powered by emerging technologies, helps fulfill the six core requirements of 6G networks.
Detailed discussions of specific integration approaches are provided in Section~\ref{Sect:Solutions}.
\color{black}

\section{Efficient bidirectional Integration Solutions for 6G and Beyond Networks} \label{Sect:Solutions}

\subsection{Transfer Learning}\label{Sect:TL}
\begin{figure}[!t]
\centering
\includegraphics[width=.81\linewidth]{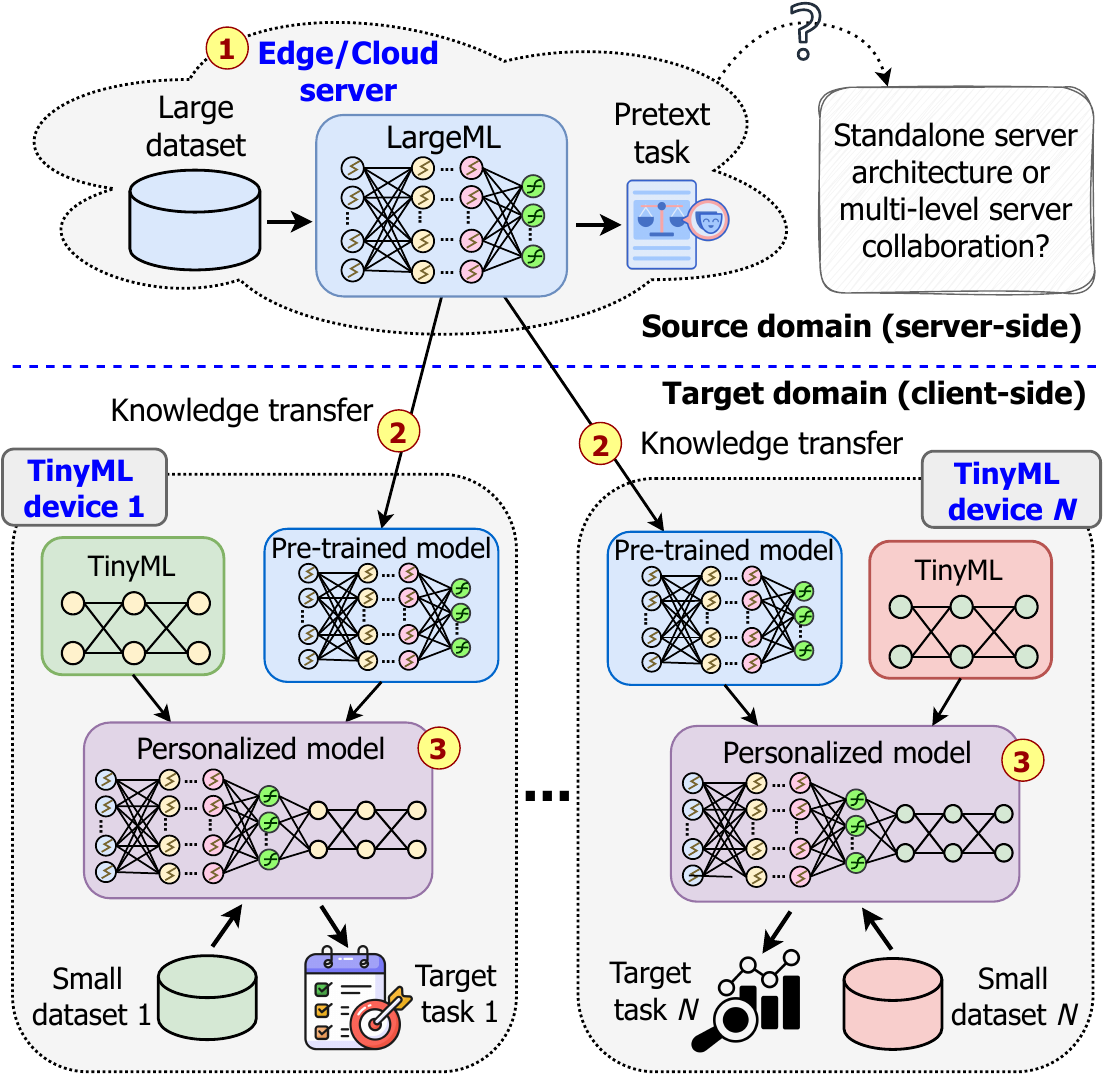}
\caption{TinyML--LargeML system with TL.}
\label{Fig:TL_sol}
\end{figure}


{TL, a subset of ML methods, enables knowledge reuse from pre-trained models across related tasks and domains (readers can find more details of this technique in \cite{wang2021introduction})}, providing an effective mechanism for TinyML--LargeML integration in 6G networks. By combining large-scale server-based training with lightweight on-device adaptation, TL supports both scalability and personalization. A generic TL framework for integration consists of three main stages (see Fig.~\ref{Fig:TL_sol}).

\circled{1} \textbf{Server-side training}:
A LargeML model $f_S$ is trained on a source-domain dataset ${{\cal D}_S}$ using high-performance cloud or edge servers. The objective is to optimize model parameters comprising weights and biases by minimizing the loss function
\begin{align}\label{eq:loss_server}
    {f_S} = \arg \mathop {\min }\limits_{f_S} \frac{1}{{{|{\cal D}_S|}}}\sum\nolimits_{i \in {{\cal D}_S}}^{} {{\cal L}\left( {{{\bf{y}}_i},{f_S}\left( {{{\bf{x}}_i}} \right)} \right)} , 
\end{align}
where ${\left\{ {{{\bf{x}}_i},{{\bf{y}}_i}} \right\}_{i \in {{\cal D}_S}}}$ represents the training samples, and
${{f_S}\left( {{{\bf{x}}_i}} \right)}$ denotes the model's activation after processing input feature ${\bf{x}}_i$.
The loss function ${\cal L}(\cdot)$ varies depending on the learning task and may include metrics such as cross-entropy, mean squared error, MAE, or log-likelihood. 

Different server-side training architectures can be adopted \cite{tao2021hybrid}, including: (\textit{i}) hybrid training (cloud pre-training with edge fine-tuning), (\textit{ii}) distributed training (workload split between cloud and edge), (\textit{iii}) hierarchical processing (edge pre-processing before cloud training), and (\textit{iv}) enhanced availability (separating edge servers during cloud outages). Compared to standalone setups (e.g., two-tiered client-edge/client-cloud), multi-level collaboration offers greater flexibility in balancing communication and computational overhead.

\circled{2} \textbf{Pre-trained knowledge transfer}:
After training in the source domain, selected model components $t_S \subset f_S$ are transferred to the target domain.
The effectiveness of this process relies on answering four guiding questions \cite{jang2019learning,pan2009survey}:
\begin{itemize}
    \item \textit{What to transfer}: The transferred knowledge may include model parameters (weights and biases), feature extractors, embeddings, or even relational structures. The choice depends on the similarity between source and target tasks. Parameters are most effective when tasks are closely related, while feature- or relational-based knowledge suits domains with shared patterns but different label spaces.
    \item \textit{When to transfer}: Knowledge can be transmitted within \cite{parsaeefard2022efficient}:
    (\textit{i}) \textit{Real-time}, immediately after model updates, enabling rapid adaptation in dynamic environments;
    (\textit{ii}) \textit{Periodic/semi-real-time}, following scheduled intervals (e.g., daily, weekly), useful in stable or predictable contexts;
    and (\textit{iii}) \textit{On-demand}, initiated by the TinyML device when additional knowledge is required, reducing communication overhead in autonomous scenarios.
    \item \textit{Where and how to transfer}: The ``where'' aspect refers to identifying the target domain where the knowledge will be applied (e.g., TinyML devices). The ``how'' aspect ensures compatibility and maximizes the effectiveness of knowledge use in the target domain.
    Effective transfer depends on both domain similarity and the selection of optimization techniques. Transfer is typically more successful between domains with overlapping characteristics.
\end{itemize}

Examined through these guiding questions, TL can be categorized according to several criteria \cite{pan2009survey}:
\begin{itemize}
    \item \textit{Label availability}: Transductive (labels only in source), inductive (labels only in target), or unsupervised (no labels in either domain).
    \item \textit{Domain alignment}: Homogeneous (shared feature and label spaces) and non-homogeneous (otherwise).
    \item \textit{Transfer mechanism}: It includes approaches: 
    (\textit{i}) \textit{Instance-based TL}, re-weighting or selecting relevant source samples;
    (\textit{ii}) \textit{Feature-based TL}, learning shared representations via dimensionality reduction or transformations;
    (\textit{iii}) \textit{Parameter-based TL}, reusing pre-trained weights and biases as initialization for target training;
    and
    (\textit{iv}) \textit{Relational-based TL}, transferring structural knowledge such as graphs, relationships, or relational patterns.
\end{itemize}

\circled{3} \textbf{Model personalization and on-device adaptation}:
Each TinyML device $n \in {\cal N} = \left\{ {1, \ldots ,N} \right\}$ adapts the transferred model using its local dataset ${{\cal D}_n}$.
Fine-tuning typically involves freezing pre-trained layers $t_S$ while updating TinyML lightweight layers $g_n$ for device-specific tasks. 
Domain adaptation methods, such as adaptation layers \cite{ghifary2014domain} or correlation alignment (CORAL) \cite{sun2016deep},
help reduce distributional shift between ${\cal D}_S$ and ${\cal D}_n$ such that
\begin{align}\label{eq:loss_coral}
    {g_n} = \arg \mathop {\min }\limits_{{g_n}} \frac{1}{{{|{\cal D}_n|}}}\sum\limits_{j \in {{\cal D}_n}}^{} {{\cal L}\left( {{\bf{y}}_j^n,{t_S}\left( {{\bf{x}}_j^n} \right)} \right)}  + \lambda {\ell _{{{\cal D}_S},{{\cal D}_n}}},
\end{align}
where \(\lambda>0\) controls the trade‑off, and \(\ell_{\mathcal D_S,\mathcal D_n}\) quantifies the discrepancy between domains, enabling personalized learning while keeping all sensitive data on‑device.

Recent studies have explored TinyML--LargeML integration via TL across diverse applications. 
For example, {\normalfont\scshape TinyTL} \cite{cai2020tinytl} uses ProxylessNAS-Mobile for source-domain pre-training and fine-tunes TinyML devices with a once-for-all network, applying parameter-based TL for tasks like object and facial attribute recognition of cars, aircraft, flowers, birds, pets, food, and celebrities.
Similarly, \cite{ostrovan2022tinyml} introduces a bidirectional integration using CNN pre-training with TensorFlow Lite Micro, pruning, and quantization for in-car presence detection and classification. 
In \cite{profentzas2022microtl}, a bidirectional integration system with feature-based TL leverages a deep neural network (DNN) source-domain backbone and a CMSIS-NN target model for IoT health trackers, supporting object and activity recognition.
Furthermore, \cite{azevedo2023detecting} applies feature-based TL with VGG-16, DenseNet, and MobileNet backbones, deployed through Edge Impulse's EON Tuner for face mask detection.
Another study \cite{hayajneh2023tiny} combines a DNN-long short-term memory (LSTM) model with parameter-based TL for soil humidity prediction in smart agriculture, fine-tuned on TinyML devices with TensorFlow Lite Micro and quantization.
In \cite{hayajneh2024tinyml}, the authors introduce
a feature-based TL bidirectional system for motion classification, using principal component analysis (PCA) for dimensionality reduction and a CNN-LSTM trained on large datasets, with Raspberry Pi as an edge intermediary platform.
Finally, \cite{kwon2024tinytrain} presents a feature-based TL framework enhanced with meta-learning, using pre-trained MCUNet, MobileNetV2, and ProxylessNAS, then fine-tuned on TinyML devices via a task-adaptive sparse-update method to support few-shot cross-domain classification under dynamic constraints.

\begin{figure}[!t]
\centering
\includegraphics[width=.81\linewidth]{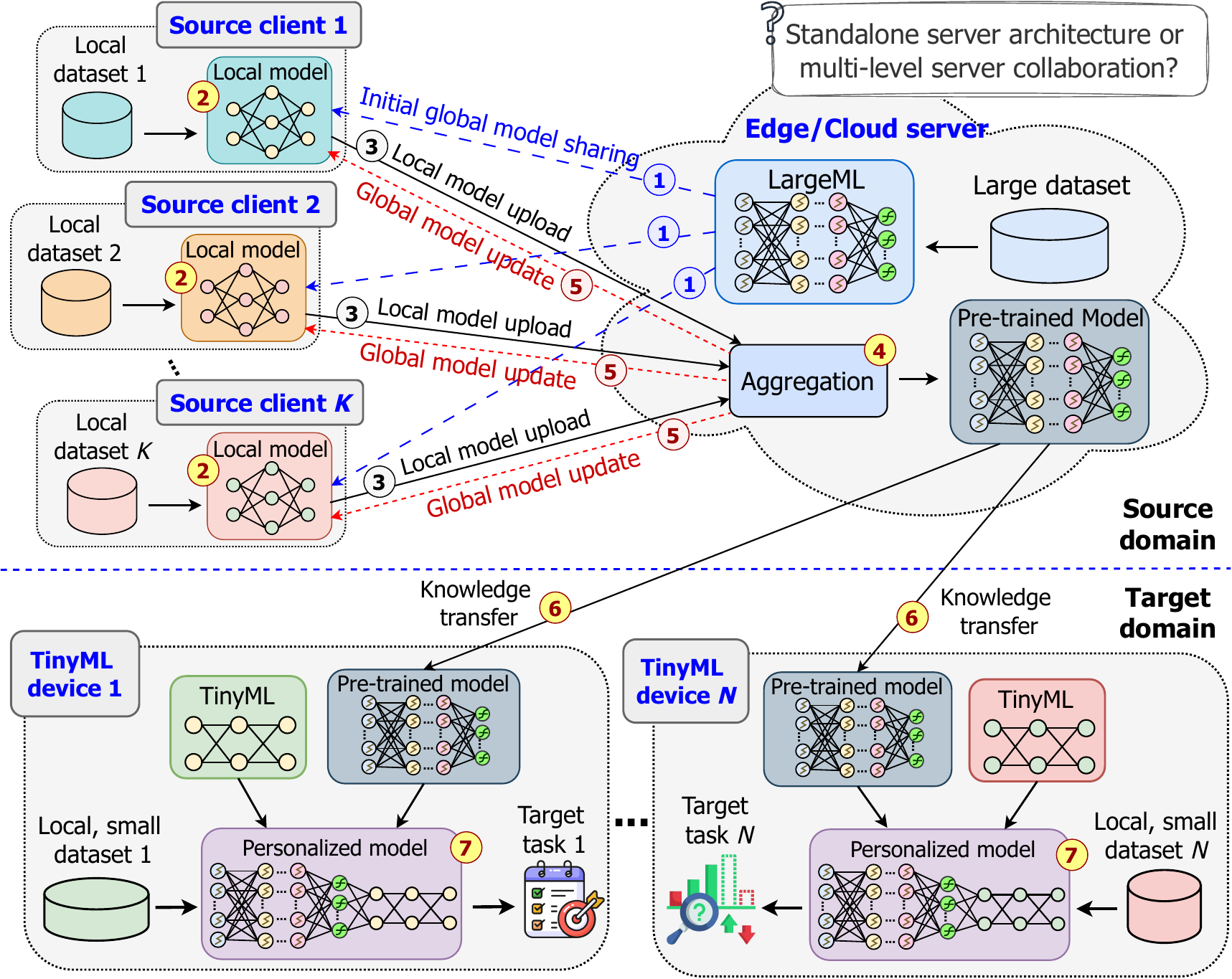}
\caption{TinyML--LargeML system with FTL.}
\label{Fig:FTL}
\end{figure}

\subsection{Federated Transfer Learning (FTL)}\label{Sect:Advanced_TL}

{TL methods (Section~\ref{Sect:TL}) work only when source and target domains share related tasks and some dataset similarities. In 6G systems, these conditions rarely hold because massive device populations generate highly non‑IID, heterogeneous, and context‑dependent data. FTL overcomes this by combining FL's collaborative training with TL's cross‑domain adaptation: source‑domain clients train the initial global model, and target‑domain clients then adapt and personalize it to their own data and tasks (see \cite{yang2019federated} for more details).}

\subsubsection{Vanilla Federated Transfer Learning}
{Vanilla FTL is the core form of FTL in which multiple parties jointly train a model without sharing raw data, even when their feature spaces and data distributions differ. It links local feature extractors to a shared prediction model optimized through privacy‑preserving gradient exchange. This separation supports scalable knowledge reuse and fine‑grained personalization across heterogeneous 6G devices, making it well suited for privacy‑sensitive TinyML–LargeML deployments.}


Fig.~\ref{Fig:FTL} illustrates the workflow of this FTL-based approach. The FTL process begins with the standard FL protocol, differing primarily in the initialization of the global model. Specifically, the FL phase includes the following steps.
    \begin{itemize}
        \item[\circled{1}] \textbf{Initial global model sharing}: Unlike conventional FL, where the global model starts with random weights, FTL initializes it with a pre-trained LargeML model $f_S$, trained on the server's large dataset ${\cal D}_S$ such that
        $f_S^{\left( 0 \right)} = \arg \mathop {\min }\limits_{{f_S}} \frac{1}{{{|{\cal D}_S|}}}\sum\nolimits_{i \in {{\cal D}_S}}^{} {{\cal L}\left( {{{\bf{y}}_i},{f_S}\left( {{{\bf{x}}_i}} \right)} \right)}$.       
        Once trained, $f_S^{\left( 0 \right)}$ is distributed to a source client set ${\cal K} = \left\{ {1, \ldots ,K} \right\}$.         
        \item[\circled{2}] \textbf{Local training}: In round $t$, each source client ${k \in {\cal K}}$ exploits its dataset ${{\cal D}_k}$ to train a local copy of the model using the loss function $f_{S,k}^{\left( t \right)} = \arg \mathop {\min }\nolimits_{f_S^{\left( {t - 1} \right)}} \frac{1}{|{\cal D}_k|}\sum\nolimits_{j \in {{\cal D}_k}}$ $ {\cal L}\big( {{\bf{y}}_j^k,f_S^{\left( {t - 1} \right)}\left( {{\bf{x}}_j^k} \right)} \big)$, adapting the model to local patterns.
        \item[\circled{3}] \textbf{Local model update}: Subsequently, each client uploads its updated parameters $f_{S,k}^{\left( t \right)}$ to the server for aggregation. 
        \item[\circled{4}] \textbf{Server-side aggregation}: The server aggregates client updates into a new global model using either FedAvg \cite{mcmahan2017communication}, proportion‑based weighting \cite{zhao2021federated}, or advanced heterogeneity‑aware aggregation methods to minimize the global loss:
        $f_S^{\left( t \right)} = \arg \mathop {\min }_{f_{S,k}^{\left( t \right)}} \sum\nolimits_{k \in {\cal K}}^{} \frac{r_k}{|{\cal D}_k|}\sum\nolimits_{j \in {{\cal D}_k}}$ ${{\cal L}\big( {{\bf{y}}_j^k,f_{S,k}^{\left( t \right)}\left( {{\bf{x}}_j^k} \right)} \big)} $,
        where ${r_k} = {|{\cal D}_k|}/\sum\nolimits_{k = 1}^K {{|{\cal D}_k|}} $ is the ratio of client $k$'s data size to the total data size. 
        %
        \item[\circled{5}] \textbf{Global model update}: The updated global model is then sent back to the source clients.
    \end{itemize}
    Steps~\circled{2} to \circled{5} are repeated until the global model reaches satisfactory performance.

Step \circled{6} presents a \textbf{knowledge transfer} phase.
Once the global model has been sufficiently trained and demonstrates strong generalization, the server transfers the useful components of the pre-trained global model $f_S$, denoted $t_S \subset f_S$, to the target TinyML devices (target clients). This transferred model encapsulates the knowledge learned from source clients.

Step \circled{7} stands for \textbf{model personalization and on-device adaptation}.
Each target TinyML device $n \in {\cal N} = \left\{ {1, \ldots ,N} \right\}$ fine-tunes $t_S$ with its smaller, domain-specific dataset ${\cal D}_n$.
Depending on task similarity, adaptation may involve updating all layers $t_S$ or only a subset.
TinyML devices employ lightweight models $g_n$, trained with the objective function \eqref{eq:loss_coral}.
This enables efficient personalization without retraining from scratch, conserving computational resources and reducing latency.

Recent studies have applied FTL to integrate TinyML and LargeML across diverse domains.
For example, {\normalfont \scshape TinyFedTL} \cite{kopparapu2022tinyfedtl}, a parameter-based FTL framework, pre-trains MobileNetV2 and adapts Perf-MobileNet (part of the TinyML Perf Benchmark) for personalized model adaptation on IoT edge devices, improving the quality of experience (QoE) under resource constraints without compromising privacy.
{\normalfont \scshape FedHealth} \cite{chen2020fedhealth}, a parameter-based FTL framework, employs a CNN backbone and fine-tuning for wearable healthcare, supporting activity recognition and auxiliary Parkinson's disease diagnosis with high accuracy and privacy preservation.
ACGAN-FTL \cite{guo2024federated} integrates auxiliary classifier generative adversarial
networks for Raspberry Pi-based TinyML devices, predicting the quality of pre-baked carbon anodes in industrial production.
In \cite{ficco2024federated}, a feature-based FTL bidirectional integration system using Genann outperformed standalone FL, TL, and TensorFlow Lite fusion across classification and regression tasks, demonstrating effective device-specific adaptation on Arduino WiFi Rev2, MKR1010, ESP8266, and ESP32.

From a hierarchical FTL perspective, {\normalfont \scshape IoTDefender} \cite{fan2020iotdefender} employs three edge servers running FL, aggregated at a cloud server, with CNN-based pre-training and parameter-based TL for intrusion detection in resource-constrained IoT networks.
Upon this, HFTL \cite{putra2023hftl} introduces a hierarchical FTL framework that incorporates additional fog servers. These fog servers aggregate models from associated edge servers before forwarding them to a cloud server for further aggregation. 
Using VGG-16, ResNet50V2, and MobileNet as pre-trained models, fine-tuned with an additional DNN, HFTL achieves secure and efficient fault classification in additive manufacturing, with parameter-based TL is the core approach.

\subsubsection{Advanced Federated Transfer Learning}

Federated Meta-Learning (FML) aims to achieve a higher level of generalization and adaptability by enabling models to learn new tasks rapidly through meta-knowledge acquired from diverse clients or domains \cite{liu2023federated}. Unlike standard FTL, transferring knowledge across related tasks, FML trains models to adapt to previously unseen tasks. This ``learning how to learn" capability improves robustness and broadens applicability, supporting both knowledge transfer and efficient task-specific adaptation.

Several studies have explored FML for TinyML--LargeML integration.
{\normalfont \scshape TinyReptile} \cite{ren2023tinyreptile} combines FL and online learning to collaboratively train RNN initializations on resource-constrained IoT devices, enabling rapid adaptation to new targets. Evaluated on sine-wave, Omniglot, and keyword spotting datasets, it demonstrated efficient implementation on Raspberry Pi 4 and Cortex-M4 MCU using parameter-based TL.
{\normalfont \scshape TinyMetaFed} \cite{ren2023tinymetafed} extends this approach with an algorithm that reduces energy consumption and communication overhead, accelerates convergence, and stabilizes training on similar TinyML devices.
Building on both {\normalfont \scshape TinyReptile} and {\normalfont \scshape TinyMetaFed}, the framework in \cite{ren2024device} introduces an advanced prototype with a CNN backbone and MLPerf Tiny Benchmark, supporting personalized adaptation for handwritten character recognition, keyword spotting, and smart-building presence detection.
Finally, {\normalfont \scshape PMFed} \cite{jia2024personalized} applies a pre-trained CNN fine-tuned on the Raspberry Pi 3B for personalized adaptation in IoT health monitoring, achieving efficient performance in defibrillation support, atrial fibrillation detection, and human activity recognition.

On the other hand, Federated Foundation Models (FFMs) integrate the strengths of Foundation Models (FMs) and FL, enabling decentralized clients to collaboratively adapt pre-trained LargeML models without exchanging raw data. This approach combines the privacy advantages of FL with the TL benefits of TFMs \cite{ren2024advances}.
Federated Fine-Tuning (FedFT) of FMs allows each client to locally personalize a shared model while maintaining data confidentiality. 
To reduce computational overhead, PEFT techniques such as LoRA, adapter modules, and prompt tuning are widely applied \cite{ren2024advances}.

Recent work has advanced FFMs and FedFT for FMs on resource-constrained edge devices.
{\normalfont \scshape FedPT} \cite{gao2024fedpt} introduces a federated proxy-tuning method, reconstructing full-sized LoRA modules from predictions for server-side aggregation, reducing computational and communication overhead but struggling under non-IID data.
To address heterogeneity, {\normalfont \scshape HetLoRA} \cite{cho2024heterogeneous} standardizes LoRA modules via zero-padding and rank truncation, improving convergence in tasks such as multi-session chat dialogue and text summarization.
{\normalfont \scshape CAFF} \cite{pfeiffer2024efficient} further improves accuracy and fairness in language and vision classification tasks by combining tiny transformers with a layer-wise FedFT and NN selection strategy.
{\normalfont \scshape FedD2P} \cite{atapour2024leveraging} distills aggregated knowledge into a prompt generator, enabling frozen FMs to adapt efficiently to downstream vision tasks (e.g., image classification or satellite imagery analysis).
Moreover, 
{\normalfont \scshape FwdLLM} \cite{xu2024fwdllm} reduces footprint and accelerates convergence by using BP-free training with PEFT and injecting minor self-generated perturbations into model parameters.

To further improve efficiency in heterogeneous environments, several studies integrate sparse architectures like MoE with PEFT.
For instance, {\normalfont \scshape FedFMSL} \cite{wu2024fedfmsl} incorporates a two-stage MoE-based FedFT, training global experts before fine-tuning local experts for improved personalization,
with sparsely-activated LoRA matrices guided by historical performance tracked in a capability queue.
This approach yields performance gains and resource savings in various image classification tasks, such as food recognition, land use, land cover, and human action detection.
{\normalfont \scshape FedMoE} \cite{mei2024fedmoe} extends {\normalfont \scshape FedFMSL} by dynamically activating submodels via global expert recommendations, achieving faster convergence in NLP tasks such as text classification, summarization, and reading comprehension.
Finally, {\normalfont \scshape A$^3$SMoE} \cite{tran2025revisiting} introduces resource-conditional computation, activating sparsely-activated MoEs based on client resource availability, surpassing prior LoRA-based techniques in instruction-tuning and complements existing heterogeneous LoRA methods.

\subsection{Split Learning}
\label{Sect:SL}
SL is an ML paradigm for collaborative model training in resource-constrained environments, allowing devices to jointly train a complete model while enhancing learning efficiency (see \cite{lin2024split} for more details).
SL is particularly suitable for integrating TinyML and LargeML within 6G and future network architectures. By partitioning the model across client devices and servers, SL allows TinyML devices to participate in training and inference tasks without becoming overwhelmed by the computational demands of LargeML while also preserving data privacy.
Two main SL variants can be applied to TinyML--LargeML systems: vanilla SL (VSL) and parallel SL (PSL).

\begin{figure*}[!t]
\centering
\includegraphics[width=.81\linewidth]{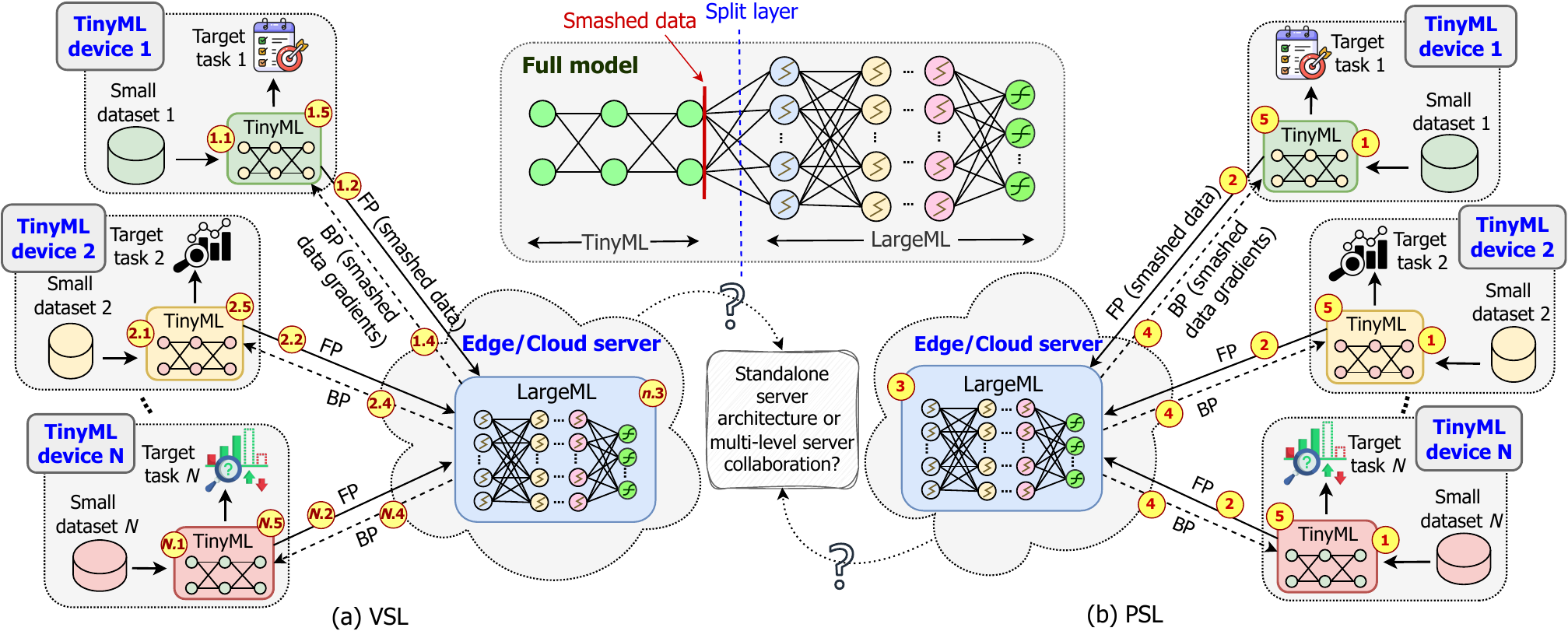}
\caption{TinyML--LargeML systems: (a) VSL and (b) PSL.}
\label{Fig:SL_sol}
\end{figure*}

\subsubsection{Vanilla Split Learning}\label{Sect:VSL}
VSL is the foundational form of SL in which the training process proceeds sequentially between clients and a coordinating server \cite{vepakomma2018split}.
The server can follow a standalone architecture (e.g., client-edge or client-cloud) or a multi-level collaborative model. 

Fig.~\ref{Fig:SL_sol}(a) illustrates a VSL-based integration of TinyML--LargeML, in which a standalone cloud or edge server collaborates with multiple TinyML devices during training.
Typically, a smaller sub-model ${\bf{T}}_n$ resides on each TinyML device, and a larger sub-model $\bf{L}$ is hosted on the server \cite{vepakomma2018split}.
Each client $n \in {\cal N}$ trains on its local dataset ${\cal D}_n = \left\{ {{\bf{x}}_j^n,{\bf{y}}_j^n} \right\}_{j = 1}^{{|{\cal D}_n|}}$.
Together, they form a temporary global model ${{\bf{G}}_n} = \left\{ {{{\bf{T}}_n};{\bf{L}}} \right\}$.
Training proceeds over $R$ outer rounds, where clients interact with the server in a round-robin fashion.
For each client, $T$ inner iterations are executed before moving to the next.
During each inner iteration, it particularly proceeds:
\begin{itemize}
    \item Step $(n.1)$: Client $n$ performs forward propagation (FP) through its local model ${\bf{T}}_n$, up to the designated split (or cut) layer. 
    The objective is to minimize its local loss function $\bar {\cal L}\left( {{{\bf{G}}_n}|{{\cal D}_n}} \right) \buildrel \Delta \over = \frac{1}{{{|{\cal D}_n|}}}\sum\nolimits_{j \in {{\cal D}_n}}^{} {{\cal L}\left( {{\bf{y}}_j^n,{{\bf{G}}_n}\left( {{\bf{x}}_j^n} \right)} \right)}$.
    This produces intermediate feature representations, called \textit{smashed data}, ${{{\bf{T}}_n}\left( {{\bf{x}}_n^{}} \right)}$.
    \item Step $(n.2)$: The smashed data and labels, $\left\{ {{{\bf{T}}_n}\left( {{\bf{x}}_n^{}} \right),{\bf{y}}_n^{}} \right\}$, are transmitted to the server. Data may be sent in full batches or mini-batches depending on optimal hyperparameters or task-specific requirements \cite{li2014efficient}.
    \item Step $(n.3)$: The server continues FP using $\bf{L}$, computes the gradients of the loss function $\bar {\cal L}\left( {{{\bf{G}}_n}|{{\cal D}_n}} \right)$ to generate $\nabla{\bar{\cal L}}({\bf{L}})$ and $\nabla{\bar{\cal L}}({\bf{T}}_n({\bf{x}}_n))$.
    Then, it performs server-side local backward propagation (BP) by updating ${\bf{L}} \leftarrow {\bf{L}} - \eta_S\nabla{\bar{\cal L}}({\bf{L}})$, where $\eta_S$ is the server learning rate.
    \item Step $(n.4)$: The server sends the smashed data gradients $\nabla{\bar{\cal L}}({\bf{T}}_n({\bf{x}}_n))$ back to client $n$.
    \item Step $(n.5)$: Client $n$ updates its local sub-model as
    ${\bf{T}}_n \leftarrow {\bf{T}}_n - \eta_n \nabla{\bar{\cal L}}({\bf{T}}_n({\bf{x}}_n))$, where $\eta_n$ is the client learning rate.
    For consistency, the terms $\eta_S$ and $\eta_n$ are reused with the same definitions in subsequent sections.    
\end{itemize}
This exchange completes one inner iteration.
During the outer iterations, the aim is to minimize the global loss function
$\mathop {\min }\limits_{{{\{\bf{G}}_n}\}_{\forall n}} 
    {{\sum\nolimits_{n \in {\cal N}}^{} {{|{\cal D}_n|}\bar {\cal L}\left( {{{\bf{G}}_n}|{{\cal D}_n}} \right)} }}\big/{{\sum\nolimits_{n \in {\cal N}}^{} {{|{\cal D}_n|}} }}$.
After convergence, TinyML devices can employ the global model ${\bf{G}}$ for inference on a designated target task.

The integration of VSL-based TinyML with LargeML has enabled a diverse range of applications for tasks in resource-constrained environments.
Specifically, a two-stage wireless channel adaptive split inference method \cite{lee2023wireless} addresses edge device memory and energy limits by dynamically adjusting split points based on real-time channel gain, improving privacy, inference accuracy, and adaptability.
Similarly, {\normalfont \scshape ARES} \cite{samikwa2022ares}, an adaptive resource-aware SL scheme, mitigates stragglers through device-specific split layers, balancing energy use, training time, and throughput variability.
An adaptive SL algorithm based on Lyapunov theory \cite{li2024adaptive} further reduces training delay and energy consumption by dynamically selecting split layers and allocating server-side resources in edge networks.

When extended to multi-level server collaboration, VSL can be deployed via multi-hop SL \cite{lin2024split}, where cloud and edge servers form a hierarchical chain.
In this setup, initial layers run on TinyML devices, intermediate layers on distributed servers, and final layers on a central server \cite{wang2021hivemind}.
During training, inference proceeds forward through the network while BP sends gradients backward to update each node's parameters \cite{tirana2022role,cao2024learning}. With $N$ TinyML devices and $K$ servers, each inner iteration requires $2NK$ inference and BP steps, highlighting the need for latency‑efficient algorithms.

Beyond serial data flow, centralized smashed data routing \cite{lin2024split} can also be implemented in 6G networks by integrating ad-hoc data routing and centralized model partitioning, enabled through software-defined networking.
This approach must consider bandwidth, memory, and computing constraints. 
Effective server coordination is crucial for compatibility between model segments, especially with non-IID data distribution across TinyML devices, necessitating periodic synchronization of model parameters for consistency.

\subsubsection{Parallel Split Learning}\label{Sect:PSL}
The sequential training process of VSL leads to significant training latency and struggles to scale with more TinyML devices, particularly with the anticipated connectivity in 6G networks. Additionally, the non-IID nature of client data can bias model performance. PSL addresses these issues by allowing multiple TinyML devices to train their models concurrently, which reduces training time \cite{lyu2023optimal}, offers ``semi-scalability'' with more devices, and helps mitigate the effects of non-IID data in 6G environments.

Fig.~\ref{Fig:SL_sol}(b) illustrates the workflow of PSL-based TinyML--LargeML systems under a standalone server architecture.
Unlike VSL-based bidirectional integration, which requires $N$ sequential rounds of forward and backward propagation, PSL completes training with a single round, regardless of $N$. This is the basis of semi-scalability, meaning the number of propagation rounds is scalable on the client side. However, the server-side training workload (i.e., smashed data gradients and computational workload), still scales proportionally with $N$.
The inner-iteration training process of a PSL-based TinyML--LargeML system proceeds as follows:
\begin{itemize}
    \item[\circled{1}] \textbf{Simultaneous client-side local FP}: Each TinyML device $n \in {\cal N}$ trains on its local dataset ${{\cal D}_n}$ using its model segment ${\bf{T}}_n$, generating smashed data ${\bf{T}}_n({\bf{x}}_n)$ at the designated split layer.
    \item[\circled{2}] \textbf{FP of smashed data}: Each device transmits its smashed data and 
    labels, $\left\{ {{{\bf{T}}_n}\left( {{\bf{x}}_n^{}} \right),{\bf{y}}_n^{}} \right\}$, to the server. The server receives these inputs concurrently from all devices.
    \item[\circled{3}] \textbf{Server-side local FP and BP}: The server trains the concatenated smashed data ${\bf{S}} = \left[ {{{\bf{T}}_1}\left( {{{\bf{x}}_1}} \right); \ldots ;{{\bf{T}}_N}\left( {{{\bf{x}}_N}} \right)} \right]$ using its model segment $\bf{L}$. 
    Concurrent training accelerates the overall local FP and generates the model output $\hat{\bf{y}} = {\bf{L}}({\bf{S}})$.
    Given the predicted result $\hat{\bf{y}}$ and the concatenated ground-truth labels ${\bf{y}} = [{\bf{y}}_1;...;{\bf{y}}_N]$, the server computes the loss value $\bar{\cal L}({\bf{y}},\hat{\bf{y}})$.
    Further process includes calculating gradients for updating $\bf{L}$, i.e., $\nabla\bar{\cal L}({\bf{L}}(\bf{S}))$ and for each set of smashed data, i.e., $\nabla\bar{\cal L}({\bf{T}}_n({\bf{x}}_n))$ before performing the local BP as ${\bf{L}} \leftarrow {\bf{L}} - \eta_S \nabla\bar{\cal L}({\bf{L}}(\bf{S}))$.
    \item[\circled{4}] \textbf{BP of smashed data gradients}: The computed gradients $\nabla\bar{\cal L}({\bf{T}}_n({\bf{x}}_n))$ are returned to each respective TinyML device in parallel.
    \item[\circled{5}] \textbf{Simultaneous client-side local BP}: Each TinyML device performs local BP and updates its model parameters, i.e., ${\bf{T}}_n \leftarrow {\bf{T}}_n - \eta_n \nabla{\bar{\cal L}}({\bf{T}}_n({\bf{x}}_n))$.
\end{itemize} 
These training steps are repeated over consecutive rounds until the model converges, i.e., minimizing the global loss function
\begin{align}\label{eq:loss_EdgeSplit}
    \mathop {\min }\limits_{{\bf{L}},{{\left\{ {{{\bf{T}}_n}} \right\}}_{\forall n}}} {\sum_{n \in {\cal N}}^{} {\sum_{j \in {{\cal D}_n}}^{} {{\cal L}\left( {{\bf{y}}_j^n,{\bf{L}}\left( {{{\bf{T}}_n}\left( {{\bf{x}}_j^n} \right)} \right)} \right)} } }\frac{1}{{\sum\nolimits_{n \in {\cal N}}^{} {{|{\cal D}_n|}} }}.
\end{align}
After convergence, each TinyML device fine-tunes the model to its specific task. However, two potential issues arise:
\begin{itemize}
    \item Although PSL helps prevent overfitting caused by aligning the end-to-end output model too closely with the distribution of the previous client, the highly non-IID nature of data across TinyML devices can reduce the convergence rate.
    This may require server or client-side aggregation strategies.
    Relevant solutions include {\normalfont \scshape EdgeSplit} \cite{zhang2024resource} and FSL.    
    \item  The server assumes the primary training workloads from $N$ TinyML devices, which does not scale as $N \to \infty$. Although cloud/edge servers are generally powerful, they could still become bottlenecks in the massive connectivity expected in 6G systems. An efficient variant, PSL (EPSL), addresses this issue \cite{lin2024efficientPSL}.
\end{itemize}
Further discussion of {\normalfont \scshape EdgeSplit} and EPSL is given in Section~\ref{Sect:APSL}; details of FSL are provided in Section~\ref{Sect:FSL}.

Recent studies have applied PSL to TinyML--LargeML systems \cite{jeon2020privacy,chopra2023adaptive,kim2023bargaining}.
A privacy-sensitive PSL framework in \cite{jeon2020privacy} enables parallel SL training across multiple resource-constrained devices, mitigating overfitting from non-IID training orders and varying dataset sizes on resource-constrained devices while preserving privacy.
{\normalfont \scshape AdaSplit} \cite{chopra2023adaptive} enhances scalability in heterogeneous, resource-constrained systems by minimizing on-device computation, optimizing server-side processing, and reducing via sparse model updates without requiring server gradients on clients.
In \cite{kim2023bargaining}, PSL-based bidirectional integration is modeled as a multiplayer bargaining problem to identify optimal split-layer strategies, balancing computation, transmission energy, training time, and data privacy, while showing robustness on non-IID datasets.

In PSL-based TinyML--LargeML systems involving multi-level server collaboration, the architecture extends beyond simple client-edge or client-cloud configurations by incorporating multiple servers or hierarchical layers of servers that collaborate during the training process.
Such setups require server orchestration and aggregation strategies, necessitating advanced PSL schemes to coordinate interactions between server layers and integrate TinyML with LargeML seamlessly.
Detailed in Section~\ref{Sect:APSL}, this class of schemes is referred to as advanced PSL, where parallel-server PSL \cite{tirana2024workflow} and multi-hop PSL \cite{tirana2024mp} serve as pioneering frameworks. These frameworks support the development of more sophisticated server orchestration and aggregation methods.

\subsubsection{Advanced Parallel Split Learning}\label{Sect:APSL}
This section introduces {\normalfont \scshape EdgeSplit} and EPSL, which are advanced PSL schemes designed to address the highly non-IID nature of data across TinyML devices. Both schemes consider model aggregation at the server side. EPSL further provides full scalability.
The details of each scheme are described below.

\begin{figure}[!t]
\centering
\includegraphics[width=.81\linewidth]{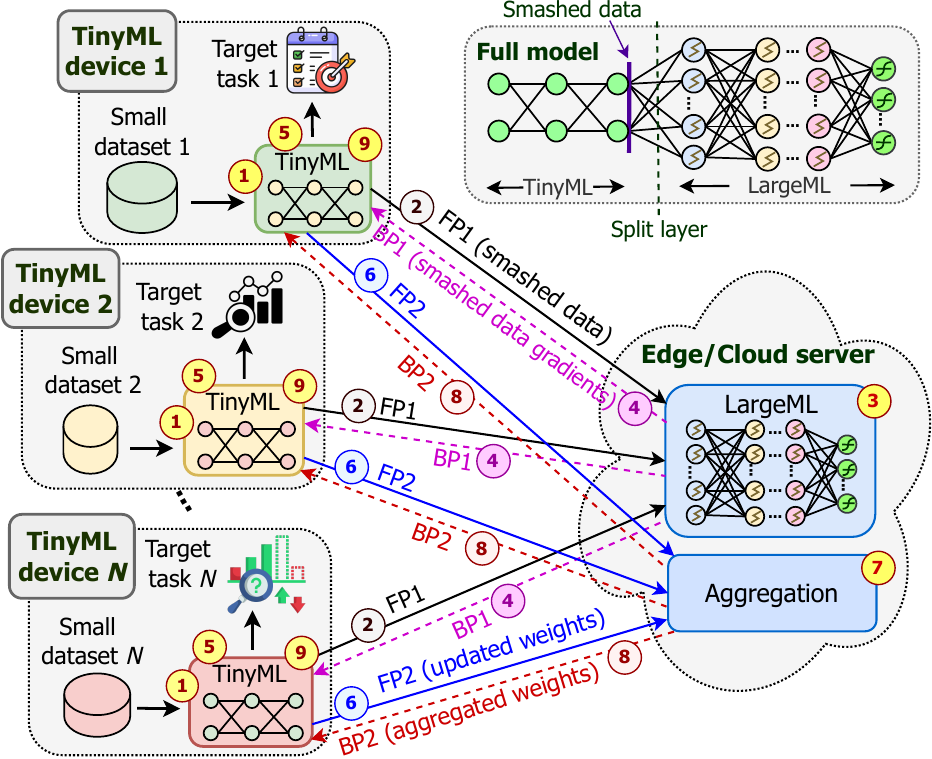}
\caption{Integrated TinyML--LargeML system with {\normalfont \scshape EdgeSplit}.}
\label{Fig:EdgeSplit}
\end{figure}

{\normalfont \scshape EdgeSplit} \cite{zhang2024resource} is an advanced PSL variant designed to address the challenge of highly non-IID data distributed across TinyML devices. Non-IID data causes local models trained on individual TinyML devices to diverge, reducing generalization and degrading global model performance. {\normalfont \scshape EdgeSplit} introduces a coordinated learning approach across diverse datasets while preserving the core benefits of PSL.
Fig.~\ref{Fig:EdgeSplit} illustrates the inner-iteration training steps of {\normalfont \scshape EdgeSplit}, which begins by a standard PSL procedure: 
\circled{1} Simultaneous client-side local FP,
\circled{2} FP1 of smashed data,
\circled{3} server-side local FP and local BP,
\circled{4} BP1 of smashed data gradients,
and
\circled{5} simultaneous client-side local BP.
The subsequent steps introduce the key advances of {\normalfont \scshape EdgeSplit} over conventional PSL.
\begin{itemize}
    \item[\circled{6}] \textbf{FP2 with updated weights}: After client-side local BP, each TinyML device forwards its updated model weights ${\bf{T}}_n$ to the server. These weights incorporate the local learning adjustments made by each device based on its non-IID data.
    \item[\circled{7}] \textbf{Server-side aggregation}: The server aggregates all updated weights received from TinyML devices using an algorithm, such as FedAvg \cite{mcmahan2017communication}:
    \begin{align}\label{eq:FedAvg}
        {{\bf{T}}_{{\rm{FedAvg}}}} = {{\sum\nolimits_{n \in {\cal N}}^{} {{|{\cal D}_n|}{{\bf{T}}_n}} }}\Big/{{\sum\nolimits_{n \in {\cal N}}^{} {{|{\cal D}_n|}} }}.
    \end{align}
    This step harmonizes learning across all devices by combining their weights and enables the server to produce a more generalized model that reflects the diversity of data distributions. 
    \item[\circled{8}] \textbf{BP2 of aggregated weights}: The server broadcasts the aggregated weights ${{\bf{T}}_{{\rm{FedAvg}}}}$ to all TinyML devices, ensuring that each device updates its model with globally informed weights that reduce the impact of non-IID data.
    \item[\circled{9}] \textbf{Updated client-side model}: 
    Each TinyML device receives the globally aggregated model ${{\bf{T}}_{{\rm{FedAvg}}}}$ and updates its weights accordingly, i.e., ${\bf{T}}_n\leftarrow {{\bf{T}}_{{\rm{FedAvg}}}}$, $\forall n$, to synchronize all devices with a consistent model.
\end{itemize}
This iterative process continues until convergence, guided by the global objective function defined in \eqref{eq:loss_EdgeSplit}.
During the inference phase, each TinyML device adapts the converged model to its specific target task.

{\normalfont \scshape EdgeSplit} is used for selected TinyML devices. In the context of widespread 6G connectivity, device heterogeneity is common, leading to variations in computing power, data distribution, and channel conditions. Cluster-based PSL (CPSL) \cite{wu2023split} addresses this by clustering devices based on shared characteristics like computing capabilities. This results in clusters with stable data distributions, enabling more efficient training and enhanced model performance.

\begin{figure}[!t]
\centering
\includegraphics[width=.81\linewidth]{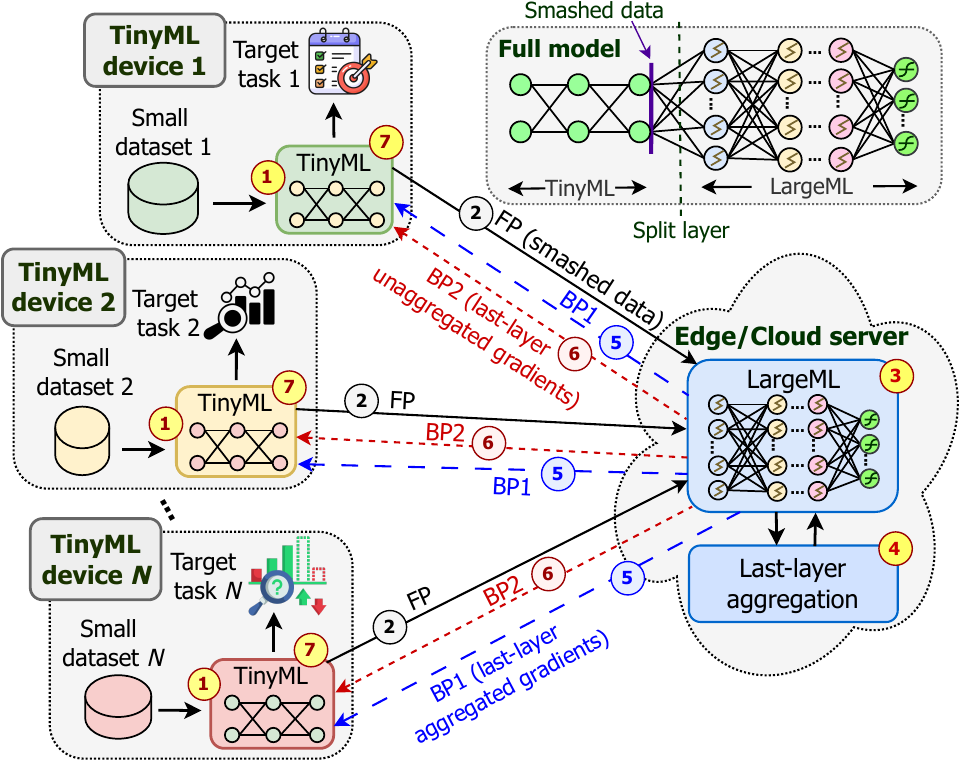}
\caption{Integrated TinyML--LargeML system with EPSL.}
\label{Fig:EPSL}
\end{figure}

Conventional PSL methods like {\normalfont \scshape EdgeSplit} and {\normalfont \scshape CPSL} scale well on the client side but still leave heavy computation on the server. EPSL \cite{lin2024efficientPSL} addresses this imbalance with a revised aggregation strategy that lowers server‑side cost and training latency, achieving full end‑to‑end scalability. Its main steps are shown in Fig.~\ref{Fig:EPSL} and summarized below.
\begin{itemize}
    \item[\circled{1}] \textbf{Simultaneous client-side local FP}: Each TinyML device $n \in {\cal N}$ independently trains on its local dataset ${{\cal D}_n}$ using its local model segment, ${\bf{T}}_n$. The device generates smashed data ${\bf{T}}_n({\bf{x}}_n)$ at the designated split layer. This step is executed in parallel across all devices.
    \item[\circled{2}] \textbf{FP of smashed data}: Devices transmit their smashed data and labels, $\left\{ {{{\bf{T}}_n}\left( {{\bf{x}}_n^{}} \right),{\bf{y}}_n^{}} \right\}$, to the server. The server processes these multiple smashed data streams concurrently.
    \item[\circled{3}] \textbf{Server-side local FP}: The server concatenates the smashed data ${\bf{S}} = \left[ {{{\bf{T}}_1}\left( {{{\bf{x}}_1}} \right); \ldots ;{{\bf{T}}_N}\left( {{{\bf{x}}_N}} \right)} \right]$, then forwards this data through its model segment $\bf{L}$ to generate the predicted output $\hat{\bf{y}} = {\bf{L}}({\bf{S}})$. 
    \item[\circled{4}] \textbf{Gradient aggregation and server-side local BP}: 
    The server uses the loss \(\bar{\mathcal L}(\mathbf y,\hat{\mathbf y})\) to compute the last‑layer activation gradients \(\mathbf G_a\), where \(\mathbf y=[\mathbf y_1;\dots;\mathbf y_N]\) is the concatenated ground‑truth label vector. Aggregating these activation gradients compresses the information passed back to clients, which substantially reduces server‑side computation, lowers training latency, and improves scalability \cite{lin2024efficientPSL}. The server then computes its own model gradients, denoted \(\mathbf G_S\).
    The parameter update for the server-side model is expressed as ${\bf{L}} \leftarrow {\bf{L}} - {\eta _S}{{\bf{G}}_S}$.
    \item[\circled{5}] \textbf{BP1 (aggregated gradients)}: The server broadcasts the aggregated activation gradients ${\bf{G}}_a$, computed at the split layer by performing BP on the last-layer aggregated activation gradients, to all participating TinyML devices. 
    \item[\circled{6}] \textbf{BP2 (unaggregated activation gradients)}: To enable more precise updates, the server also sends the unaggregated activation gradients, denoted ${\bf{G}}_u$, computed at the split layer by performing BP on the last-layer aggregated activation gradients, to the respective TinyML devices. 
    \item[\circled{7}] \textbf{Client-side local BP}: Each TinyML device uses the received gradients ${\bf{G}}_a$ and ${\bf{G}}_u$ to compute its local gradient ${\bf{G}}_n$ and update its model segment ${{\bf{T}}_n} \leftarrow {{\bf{T}}_n} - {\eta _n}{{\bf{G}}_n}$.
\end{itemize}
The EPSL training process is repeated until convergence. Similarly, \eqref{eq:loss_EdgeSplit} can be used as the global objective function for the EPSL training process. During inference, each TinyML device adapts the trained model to its specific task, enabling personalized performance on its local data.
Compared to standard PSL, EPSL reduces BP computation and communication complexity from ${\mathcal{O}}(N)$ to ${\mathcal{O}}(1)$, achieving full scalability \cite{lin2024split}. EPSL also introduces a tunable aggregation ratio, denoted $r$, in the BP phase, enabling a trade-off between computational and communication efficiency while maintaining learning accuracy. When $r=0$, EPSL reverts to PSL.

From the PSL-based multi-level server collaboration framework introduced in Section~\ref{Sect:PSL}, two extensions have been proposed: parallel-server PSL and multi-hop PSL.
In parallel-server PSL \cite{tirana2024workflow}, multiple servers concurrently process model segments for resource-constrained clients, requiring orchestration management and joint optimization of client-server assignments and scheduling to minimize training time. Due to its NP-hard nature, two scalable solutions are suggested: (\textit{i}) a decomposition-based method utilizing structural symmetry, and (\textit{ii}) a lightweight alternative with low computational overhead.
Conversely, multi-hop PSL \cite{tirana2024mp} enhances traditional multi-hop SL with pipelined, parallel execution. Each locally trained TinyML device generates smashed data at a split layer and sends it to the first server, where inference and backpropagation occur sequentially, akin to multi-hop SL in Section~\ref{Sect:VSL}. After backpropagation, devices receive synchronized gradient updates for aggregation.
Unlike multi-hop SL, multi-hop PSL reduces inference and backward steps to $(2K+2)$, independent of the number of devices $N$.

\begin{figure}[!t]
\centering
\includegraphics[width=.81\linewidth]{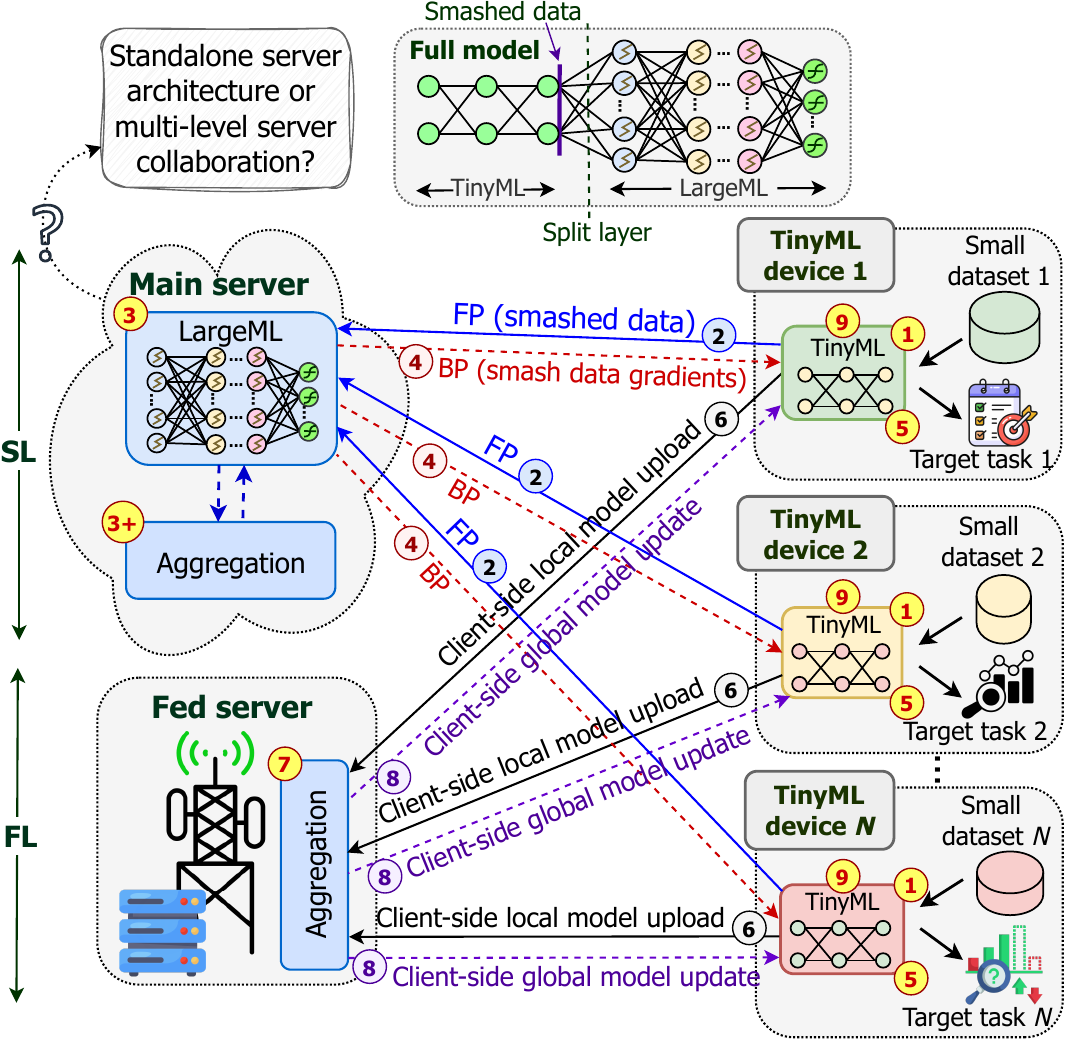}
\caption{Integrated TinyML--LargeML system with FSL.}
\label{Fig:FSL}
\end{figure}

\subsection{Federated Split Learning (FSL)}\label{Sect:FSL}


{PSL parallelizes client‑side model training (see Section~\ref{Sect:PSL}) while handling highly non‑IID client data, effectively blending the advantages of split learning and federated learning \cite{thapa2022splitfed}. FSL differs from PSL,  {\normalfont \scshape EdgeSplit}, and EPSL by distributing responsibilities across two cooperating servers: a main cloud server and a Fed server that runs FedAvg. In FSL, the Fed server aggregates updated local models from clients using FL-based techniques (see \cite{turina2021federated} for more details).}

\subsubsection{Vanilla Federated Split Learning}
{Vanilla FSL, the simplest and foundational form of FSL, merges the core ideas of SL, where the model is cut at a chosen layer and clients only compute the early part of the network, with FL, where many clients participate in training while keeping their data local.} This process is illustrated in Fig.~\ref{Fig:FSL}, with the key steps:
\begin{enumerate}
    \item[\circled{1}] \textbf{Simultaneous client-side local FP}: Each TinyML device $n \in {\cal N}$ performs local FP on its model segment, ${\bf{T}}_n$, using its local dataset ${{\cal D}_n}$ to generate smashed data ${\bf{T}}_n({\bf{x}}_n)$ at the split layer. 
    \item[\circled{2}] \textbf{FP of smashed data}: Devices concurrently transmit the smashed data and corresponding labels $\left\{ {{{\bf{T}}_n}\left( {{\bf{x}}_n^{}} \right),{\bf{y}}_n^{}} \right\}$ to the main server for further centralized processing. 
    \item[\circled{3}] \textbf{Main-server processing}: The main server processes the smashed data in both local FP and BP using either strategies of SplitFedV1 (when the server has substantial computing capacity and can efficiently execute parallel processing) or SplitFedV2 (personalized updates on the main server are essential, and client data is relatively homogeneous) \cite{thapa2022splitfed,thapa2021advancements}. 
    In SplitFedV1, each client's smashed data is separately processed and parallelized on its model segment $\bf{L}$, where
        the loss value $\bar {\cal L}( {{{\bf{y}}_n},{{{\bf{\hat y}}}_n}})$ and its gradient $\nabla \bar {\cal L}( {{{\bf{y}}_n},{{{\bf{\hat y}}}_n}})$ are calculated based on the predicted output ${{\bf{\hat y}}_n} = {\bf{L}}( {{{\bf{T}}_n}( {{{\bf{x}}_n}})})$ and ground-truth label ${{\bf{y}}_n}$. In the subsequent steps, the server aggregates the gradients of the smashed data and updates the server-side model using the FedAvg algorithm \cite{mcmahan2017communication}, expressed as
        ${\bf{L}} \leftarrow {\bf{L}} - {\eta _S}{{\sum\nolimits_{n \in {\cal N}}^{} {{|{\cal D}_n|}\nabla \bar {\cal L}\left( {{{\bf{y}}_n},{{{\bf{\hat y}}}_n}} \right)} }}\Big/{{\sum\nolimits_{n \in {\cal N}}^{} {{|{\cal D}_n|}} }}$. In SplitFedV2, the server processes client smashed data sequentially on $\bf{L}$, following a randomly determined client order represented by the list $\cal S$ and the computation of $\nabla \bar {\cal L}\left( {{{\bf{y}}_n},{{{\bf{\hat y}}}_n}} \right)$ akin to SplitFedV1. 
        Instead of aggregation, the server sequentially process the data in the order defined by $\cal S$ to update the global model after each local FP using the rule ${\bf{L}} \leftarrow {\bf{L}} - {\eta _S}\nabla \bar {\cal L}\left( {{{\bf{y}}_n},{{{\bf{\hat y}}}_n}} \right)$ and continuing until all clients in the list $\cal S$ are processed.

    \item[\circled{4}] \textbf{BP of smashed data gradients}: After performing local FP and BP on the server-side model, the main server returns the evaluation of $\nabla \bar {\cal L}\left( {{{\bf{y}}_n},{{{\bf{\hat y}}}_n}} \right)$ to TinyML devices.
    \item[\circled{5}] \textbf{Client-side local BP}: Upon receiving the gradients from the main server, each TinyML device $n \in {\cal N}$ calculates $\nabla \bar {\cal L}\left( {{{\bf{T}}_n}\left( {{{\bf{x}}_n}} \right)} \right)$ and performs local BP, updating its model as
    ${{\bf{T}}_n} \leftarrow {{\bf{T}}_n} - {\eta _n}\nabla \bar {\cal L}\left( {{{\bf{T}}_n}\left( {{{\bf{x}}_n}} \right)} \right)$.
    \item[\circled{6}] \textbf{Client-side local model upload}: Devices upload their updated local model to the Fed server for aggregation.
    \item[\circled{7}] \textbf{Fed-server aggregation}: The Fed server aggregates the client-side local models using \eqref{eq:FedAvg} to generate a global client-side model ${\bf{T}}_{\rm{FedAvg}}$ that incorporates updates from all participating TinyML devices.
    \item[\circled{8}] \textbf{Client-side global model update}: ${\bf{T}}_{\rm{FedAvg}}$ is returned to all TinyML devices. 
    
    \item[\circled{9}] \textbf{Client-side local model synchronization}: Device update their local model with the rule ${\bf{T}}_n\leftarrow {{\bf{T}}_{{\rm{FedAvg}}}}$, $\forall n$. This step ensures that all devices converge toward improved and consistent model performance.
\end{enumerate}
This FSL cycle repeats iteratively until the overall model converges. During inference, each TinyML device adapts the converged model to its specific target task.

Several recent studies have explored the use of FSL for integrating TinyML and LargeML in a range of applications \cite{han2023federated,zhu2024esfl,lin2024adaptsfl,ni2024fedsl}. Building upon SplitFedV1, {\normalfont \scshape SplitGP} supports bidirectional training for personalized client models and generalized server models, surpassing baselines \cite{han2023federated}. Similarly, ESFL enhances training stability by incorporating prior global models, using an iterative algorithm to solve its NP-hard optimization, improving efficiency over standard FL, SL, and FSL \cite{zhu2024esfl}. For SplitFedV2, {\normalfont \scshape AdaptSFL} dynamically adjusts model splitting and aggregation to optimize latency and convergence on resource-constrained devices, outperforming benchmarks \cite{lin2024adaptsfl}. Meanwhile, {\normalfont \scshape FedSL} enables robust healthcare analytics on wearables, excelling in medical imaging across diverse data \cite{ni2024fedsl}. Further, hierarchical FSL frameworks boost efficiency: one trains multiple edge server-client pairs with cloud aggregation for resilience \cite{zhang2023privacy}, while another uses parallel multi-client setups for enhanced performance on IoT devices \cite{khan2023joint}.

\subsubsection{Federated Split Models and Fine-Tuning}
Federated split FMs (FSFMs) extend the FL framework by partitioning large FMs across clients and servers, enabling distributed training and inference with reduced local resource requirements \cite{li2024synergizing}. 
Clients train or fine-tune distinct FM segments collaboratively, preserving privacy by not exposing raw data and lowering computational overhead. 
Federated split fine-tuning (FedSFT) further enables client-side personalization of model components while retaining global knowledge through coordinated updates.
Compared to standard FL and FFMs, FSFMs improve scalability by exchanging only compact smashed data rather than full model parameters \cite{li2024synergizing}.
These features position FSFMs and FedSFT as particularly suitable approaches in 6G scenarios characterized by device heterogeneity, limited computational capacity, and stringent privacy requirements.

For instance, {\normalfont \scshape SFPrompt} \cite{cao2024sfprompt} is a prompt-based FedSFT framework for privacy-sensitive, resource-constrained edge environments.
By dividing pre-trained FMs between server and client and integrating soft prompts with local data pruning, it improves FedFT efficiency while significantly reducing computation and communication costs in image classification.
{\normalfont \scshape FedSplitX} \cite{shin2023fedsplitx} introduces fine-grained FSFM with multiple model partition points to accommodate diverse client capabilities in heterogeneous, resource-constrained edge environments.
Auxiliary networks at each partition reduce latency and communication overhead, enhancing image classification accuracy.
{\normalfont \scshape FedVZ} \cite{shi2024heterogeneous} defends against gradient inversion in FSFMs using vision transformers by replacing BP with a zeroth-order optimization forward pass, enabling secure, efficient training in constrained environments.
Finally, {\normalfont \scshape PRINCE} \cite{li2025incentivizing} incentivizes high-quality device participation from multiple FSL tenants via strategic pricing, boosting FedSFT performance in tasks such as image classification, sentiment analysis, speech-to-text, and question answering.

\begin{table*}[ht!]
    \centering
    \caption{Summary of TinyML--LargeML integration solutions.}
    \begin{tabular}{|c|p{6cm}|p{4.25cm}|p{1.9cm}|p{3cm}|}
        \hline
        {\textbf{Sol.}} & 
        \centering {\textbf{Procedures}} & 
        \centering {\textbf{Characteristics}} &        
        \centering {\textbf{Representatives}} &
        \hspace{1cm} {\textbf{Advances}} \\
        \hline\hline
        {\multirow{5}{*}{\textbf{TL}}}
        & {
        {$(1)$~Server-side training 
        $\to$ $(2)$~Pre-trained knowledge transfer 
        $\to$ $(3)$~Model personalization and on-device adaptation}}
        & {- Aligns well with resource-constrained TinyML devices
        
        - Requires ideal conditions: source and target domains with distinct yet closely related datasets and tasks} &
        {{\normalfont\scshape TinyTL} \cite{cai2020tinytl}, \cite{ostrovan2022tinyml}, \cite{profentzas2022microtl}, \cite{azevedo2023detecting}, \cite{hayajneh2023tiny}, \cite{hayajneh2024tinyml}, \cite{kwon2024tinytrain}, and \cite{yilmaz2021transfer}}
        & Multi-level server collaboration: Hybrid training, distributed fashion, hierarchical processing, and enhanced availability \\
        \hline
        {\multirow{5}{*}{\textbf{FTL}} }
        & {$(1)$~Model initialization 
        $\to$ $(2)$~Source-client-side local training 
        $\to$ $(3)$~Local model upload  
        $\to$ $(4)$~Aggregation 
        $\to$ $(5)$~Global model update 
        $\to$ $(6)$~Knowledge transfer 
        $\to$ $(7)$~Model personalization and on-device adaptation

        \hspace{0.5cm}(Steps $(2)-(5)$ iterate until convergence)}
        & {- Enables collaborative learning and distinguishes between source and target clients
        
        - Combines the strengths of FL and TL, efficient in highly non-IID data across TinyML devices} &
        {{\normalfont \scshape TinyFedTL} \cite{kopparapu2022tinyfedtl}, {\normalfont \scshape FedHealth} \cite{chen2020fedhealth}, ACGAN-FTL \cite{guo2024federated}, and \cite{ficco2024federated}}
        & {- Hierarchical FTL: {\normalfont \scshape IoTDefender} \cite{fan2020iotdefender}, and HFTL \cite{putra2023hftl}
        
        - FML: \cite{ren2023tinyreptile,ren2023tinymetafed,ren2024device,jia2024personalized}

        - FFM and FedFT of FMs: \cite{gao2024fedpt,cho2024heterogeneous,pfeiffer2024efficient,atapour2024leveraging,xu2024fwdllm,wu2024fedfmsl,mei2024fedmoe,tran2025revisiting} }
        \\
        \hline
        {\multirow{5}{*}{\textbf{VSL}} }
        & 
        {$(1.1)$~First client local FP 
        $\to$ $(1.2)$~FP 
        $\to$ $(1.3)$~Server-side local FP and BP  
        $\to$ $(1.4)$~BP 
        $\to$ $(1.5)$~First client local BP 
        $\to \cdots \to$
        $(N.1)$~Last client local FP 
        $\to \cdots \to$ $(N.5)$~Last client local BP 
        $\to$ $(6)$~On-device adaptation

        \hspace{0.35cm}(Steps $(1.1)-(N.5)$ iterate until convergence)}
        & 
        {- Enables collaborative learning with sequential training among participating TinyML devices
        
        - Incurs high training latency, lacks scalability, overfits to the last trained client} &
        {{\normalfont \scshape ARES} \cite{samikwa2022ares}, frameworks in \cite{lee2023wireless} and \cite{li2024adaptive} }
        & 
        {Multi-level server collabo-ration: Multi-hop SL \cite{wang2021hivemind} and centralized ad-hoc data routing \cite{lin2024split}}
        \\
        \hline
        {\multirow{5}{*}{\textbf{PSL}}}
        & 
        {$(1)$~Simultaneous client-side local FP 
        $\to$ $(2)$~FP 
        $\to$ $(3)$~Server-side local FP and BP  
        $\to$ $(4)$~BP 
        $\to$ $(5)$~Simultaneous client-side local BP 
        $\to$ $(6)$~On-device adaptation

        \hspace{0.5cm}(Steps $(1)-(5)$ iterate until convergence)}
        & 
        {- Enables collaborative learning with parallel training among TinyML devices
        
        - Significantly reduces training latency, offers ``semi-scalability", and partially addresses non-IID data issues} &
        {{\normalfont \scshape AdaSplit} \cite{chopra2023adaptive}, frameworks in \cite{jeon2020privacy} and \cite{kim2023bargaining}}
        & {- Multi-level server collaboration: Parallel-server PSL \cite{tirana2024workflow} and multi-hop PSL \cite{tirana2024mp}
        
        - Advanced PSL: {\normalfont \scshape {EdgeSplit}} \cite{zhang2024resource} and EPSL \cite{lin2024efficientPSL}
        }\\
        \hline
        {\multirow{6}{*}{\makecell{{\normalfont \scshape \textbf{Edge}}\\   
        {\normalfont \scshape \textbf{Split}}}}}
        & 
        {$(1)$~Simultaneous client-side local FP 
        $\to$ $(2)$~FP1 
        $\to$ $(3)$~Server-side local FP and BP  
        $\to$ $(4)$~BP1 
        $\to$ $(5)$~Client-side local BP 
        $\to$ $(6)$~FP2 
        $\to$ $(7)$~Server-side aggregation 
        $\to$ $(8)$~BP2
        $\to$ $(9)$~Updated client-side model
        $\to$ $(10)$~On-device adaptation

        \hspace{0.5cm}(Steps $(1)-(9)$ iterate until convergence)}
        & 
        {- An advanced variant of PSL with additional server-side aggregation
        
        - Improves convergence rate in highly non-IID data environments while maintaining PSL benefits} &
        {\cite{zhang2024resource}}
        & {CPSL \cite{wu2023split} for clustering multiple groups of TinyML devices}\\
        \hline
        {\multirow{5}{*}{{\textbf{EPSL}}}} 
        & 
        {$(1)$~Simultaneous client-side local FP 
        $\to$ $(2)$~FP 
        $\to$ $(3)$~Server-side local FP  
        $\to$ $(4)$~Last-layer gradient aggregation and BP
        $\to$ $(5)$~BP1 
        $\to$ $(6)$~BP2 
        $\to$ $(7)$~Client-side local BP
        $\to$ $(8)$~On-device adaptation

        \hspace{0.5cm}(Steps $(1)-(7)$ iterate until convergence)}
        &
        {- An advanced variant of PSL with modified server-side aggregation
        
        - Maintains PSL benefits, reduces computational workload and training latency, offers full scalability}
        & {\cite{lin2024efficientPSL}} 
        & {} \\
        \hline
        \multirow{13}{*}{\textbf{FSL}} & 
        {The following procedures are for both \textbf{SplitFedV1} and \textbf{SplitFedV2}, except that step $(3+)$ is specified to SplitFedV1:
        
        $(1)$~Simultaneous client-side local FP 
        $\to$ $(2)$~FP 
        $\to$ $(3)$~Main-server-side local FP  
        \textit{$\to$ $(3+)$~Main-server-side aggregation and local BP (only for SplitFedV1)}
        $\to$ $(4)$~BP
        $\to$ $(5)$~Client-side local BP 
        $\to$ $(6)$~Client-side local model upload
        $\to$ $(7)$~Fed-server aggregation
        $\to$ $(8)$~Client-side global model update
        $\to$ $(9)$~Client-side local update
        $\to$ $(10)$~On-device adaptation

        \hspace{0.5cm}(Steps $(1)-(9)$ iterate until convergence)} 
        &
        {- \textbf{SplitFedV1}: An advanced variant of PSL, with aggregation on both server and client sides through main and Fed servers;  Combines the strengths of FL and SL, improves convergence rate in highly non-IID data environments
        
        - \textbf{SplitFedV2}: An advanced variant of PSL with client-side aggregation through Fed server support; Combines the strengths of FL and SL, improves convergence in highly non-IID data environments}
         &
        {{\normalfont \scshape AdaptSFL} \cite{lin2024adaptsfl},
        {\normalfont \scshape FedSL} \cite{ni2024fedsl},
        {\normalfont \scshape SplitGP} \cite{han2023federated}, and
        ESFL \cite{zhu2024esfl}
        }
        & 
        {- Hierarchical FSL: Frameworks in \cite{zhang2023privacy} and \cite{khan2023joint}
        
        - FSFMs and FedSFT of FMs: {\normalfont \scshape SFPrompt} \cite{cao2024sfprompt}, {\normalfont \scshape FedSplitX} \cite{shin2023fedsplitx}, {\normalfont \scshape FedVZ} \cite{shi2024heterogeneous}, and {\normalfont \scshape PRINCE} \cite{li2025incentivizing} } \\
                
        \hline

        \hline
        
    \end{tabular}
    \label{Tab:Comparison_solutions}
\end{table*}



\subsection{Discussions and Lessons Learned}

While Table~\ref{Tab:Comparison_solutions} outlines their key features and advances, lessons learned from the TinyML–LargeML integration provide valuable insights for future research.

First, TL-based integration (Section~\ref{Sect:TL}) fine-tunes pre-trained models on TinyML devices by updating only a subset of parameters, reducing computational overhead and enabling on-device adaptation. However, it assumes closely related source and target domains, which limits generalizability.

Second, FTL-based integration (Section~\ref{Sect:Advanced_TL}) enables client-level personalization under highly non-IID data, aligning well with 6G ubiquitous connectivity. FML further enhances adaptability through meta-learning, while FFMs combine FMs and FL to support collaborative, privacy-preserving, and decentralized FedFT; efficiency is improved via PEFT, prompt tuning, and MoE to handle model heterogeneity.

Third, SL-based integration (Section~\ref{Sect:SL}) allows resource-constrained TinyML devices to participate in full model training, but VSL suffers from high latency, limited scalability, and overfitting risks due to sequential training. PSL mitigates these issues through parallelization and partial non-IID handling.
Building on PSL, {\normalfont \scshape EdgeSplit} improves convergence under highly non-IID settings, while EPSL further reduces server-side workload and training latency, achieving full scalability

Finally, FSL-based integration (Section~\ref{Sect:FSL}) combines SL and FL to enable flexible aggregation and improved convergence under highly non-IID data, with SplitFedV1/V2 serving as foundational designs. 
Extensions such as FSFMs and FedFT of split FMs support scalable, privacy-preserving personalization via lightweight smashed-data exchange, though their SL-based architectures remain sensitive to communication latency and network conditions.

%

Notably, interdisciplinary collaboration is continued to advance and refine TinyML--LargeML integration strategies.
Active industry participation is equally important to ensure that proposed solutions are both innovative and practical for real-world deployment. 
Although 6G has not yet been officially standardized and it remains uncertain whether these approaches will fully satisfy the requirements of 6G and future networks, they represent meaningful progress toward next-generation intelligent wireless networks.
\color{black}

\section{Applications of TinyML--LargeML Integration}
\label{Sec:Application}
\subsection{Data Privacy and Network Security}\label{SecV:security-privacy}

{\textit{Toward Network Privacy}: Model inversion attacks pose major risks when handling sensitive data, and integrating TinyML with LargeML provides a privacy‑preserving architecture by keeping inference and feature extraction on-device while sending only abstracted representations or model updates to the cloud \cite{abadade2023tinyml,yang2024privacy}. TinyML supports established techniques such as federated learning \cite{abadade2023tinyml} and differential privacy \cite{villegas2024optimizing}, while its combination with LargeML enables stronger ecosystem-wide protection through FTL \cite{yang2020fedsteg}, FML \cite{liu2023federated}, and FSL \cite{thapa2022splitfed} (see Section~\ref{Sect:Solutions}). This integration is especially important in healthcare \cite{profentzas2022microtl,chen2020fedhealth,jia2024personalized,ni2024fedsl}, where TinyML wearables process physiological signals locally and LargeML servers use aggregated insights for diagnosis and prediction; similar benefits appear in smart agriculture \cite{hayajneh2023tiny}, where on-site analysis protects proprietary practices while cloud models improve forecasting, and in smart homes and cities \cite{ren2024device}, where local inference preserves user privacy while LargeML enhances personalization and resource optimization.}



\begin{figure}[!t]
\centering
\includegraphics[width=.81\linewidth]{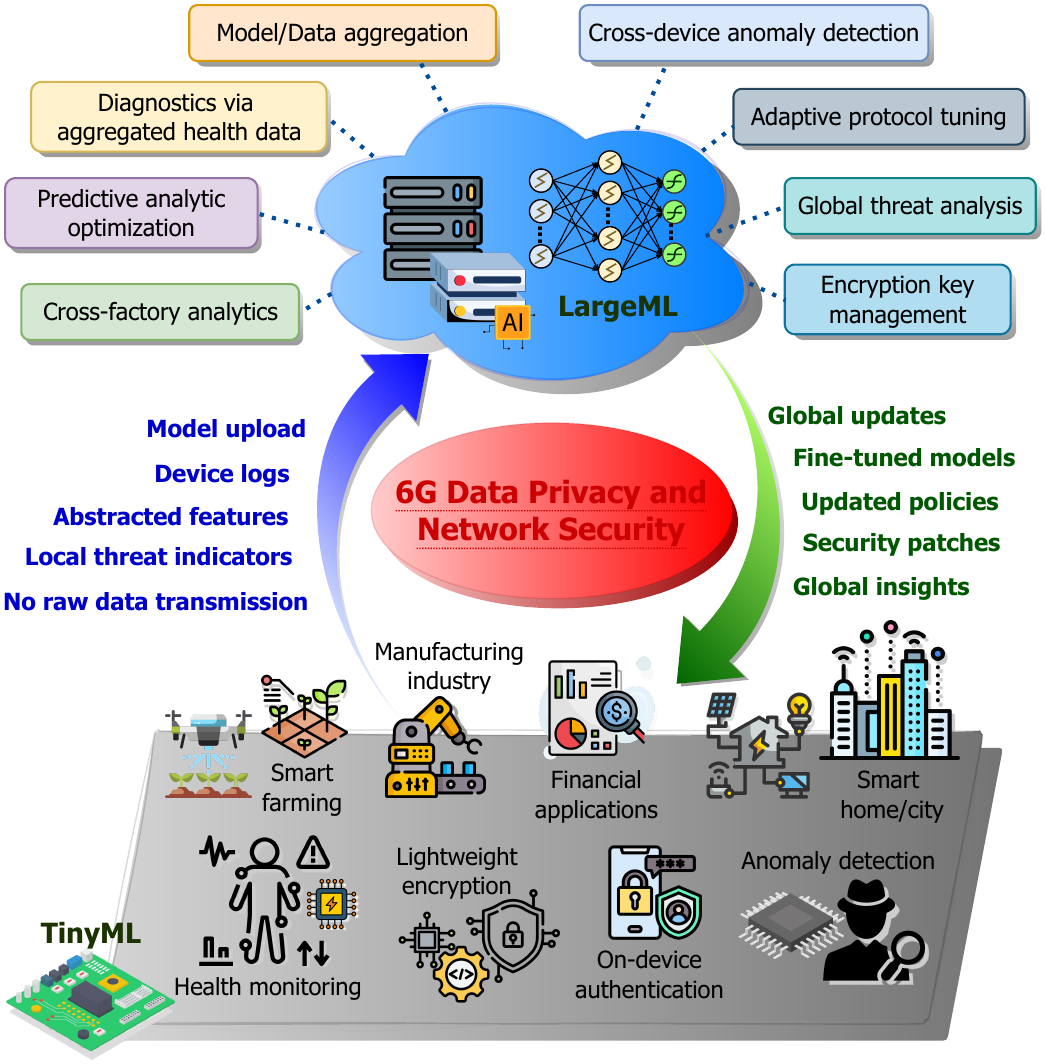}
\caption{Complementary roles of TinyML and LargeML in enhancing data privacy and network security in 6G systems.}
\label{Fig:App_Privacy_Security}
\end{figure}

{\textit{Toward Network Security}: TinyML devices operate under strict memory and compute limits, often two to three orders of magnitude below typical IoT or edge hardware \cite{lin2024efficient,lin2022device,abadade2023tinyml}, making conventional security mechanisms impractical, yet their integration with LargeML enables a multi-layered 6G defense in which TinyML performs lightweight, real‑time anomaly detection at the edge \cite{arcot2024tinyml} while LargeML correlates traffic patterns across devices to uncover complex attacks \cite{yilmaz2021transfer}, as demonstrated by {\normalfont \scshape IoTDefender} \cite{fan2020iotdefender}, which applies FTL across edge servers and cloud CNNs for intrusion detection under hardware constraints. To secure communication, TinyML supports lightweight encryption on-device \cite{patil2024securing}, while LargeML manages key distribution and protocol orchestration, dynamically updating keys to reinforce communication channels \cite{buban2025encrypted}. This hybrid approach balances strong security with device limitations and underpins the 6G privacy‑and‑security ecosystem illustrated in Fig.~\ref{Fig:App_Privacy_Security}.}

\subsection{Network Management}\label{SecV:network-management}
The rise of IoT devices and mobile users has pushed 5G networks to support ultra-reliable low-latency communication (uRLLC), massive machine-type communication (mMTC), and enhanced mobile broadband (eMBB) using network slicing, software-defined networking (SDN), network function virtualization (NFV), and multi-access edge computing (MEC) \cite{benzaid2020ai}.
Here, network slicing creates virtual networks on shared infrastructure, SDN and NFV provide flexibility through decoupling and virtualization, and MEC reduces latency via edge processing. As 6G moves toward user-centric and intelligent management, these traditional mechanisms become insufficient, necessitating closed-loop automation powered by ML and big data analytics \cite{chen2024big}.

\begin{figure}[!t] 
\centering
\includegraphics[width=.98\linewidth]{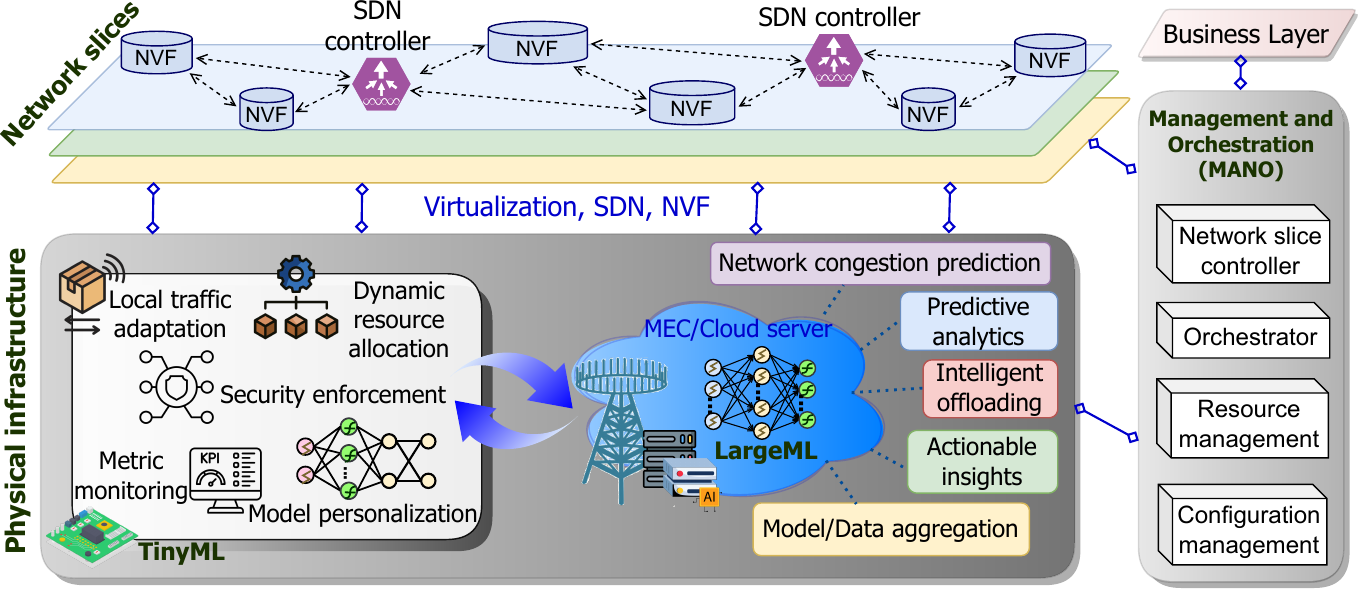}
\caption{Applications of TinyML--LargeML integration in 6G network slicing, SDN, NFV, and MEC management.}
\label{Fig:App_NetworkManagement}
\end{figure}

TinyML--LargeML integration offers a transformative approach by combining edge efficiency with large-scale analytics for adaptive management. TinyML enables localized optimization by monitoring metrics, like latency and traffic, to sustain low-latency quality-of-service (QoS) in applications such as smart factories \cite{soro2021tinyml}. 
LargeML aggregates data across nodes, detects complex patterns, predicts failures, and orchestrates network-wide resources, for instance, rerouting traffic to preserve uRLLC, mMTC, and eMBB \cite{aouedi2024deep}. 
A notable example is a Lyapunov-based adaptive split learning framework that dynamically assigns split layers to TinyML devices, reducing latency and energy consumption \cite{li2024adaptive}. Fig.~\ref{Fig:App_NetworkManagement} illustrates such bidirectional integration across domains.
Applications span major technologies:
(\textit{i}) In \textit{network slicing}, TinyML monitors metrics for real-time resource reallocation, while LargeML predicts congestion and optimizes slice provisioning \cite{khan2024edge,bega2020network};
(\textit{ii}) In \textit{SDN}, TinyML detects anomalies (e.g., DDoS attacks) at the edge, and LargeML enhances traffic control by predicting congestion and optimizing routes \cite{latah2019artificial,ali2024ddos};
(\textit{iii}) In \textit{NFV}, TinyML mitigates performance degradation, while LargeML automates resource scaling \cite{nekovee2020towards};
(\textit{iv}) In \textit{MEC}, TinyML minimizes latency through on-device inference, complemented by LargeML's orchestration to ensure efficient, low-latency operation \cite{cao2019intelligent}. 
Examples include {\normalfont \scshape SplitGP} \cite{han2023federated}, an FSL strategy that balances personalization (on-device TinyML) with generalization (MEC-side LargeML), enabling robust training-inference optimization across diverse inputs.

\subsection{Intent-based Networking}\label{SecV:IBN}
Intent-based networking (IBN) is a management paradigm that translates high-level business objectives, or human-readable ``intents'', into automated network configurations \cite{clemm2022rfc}.
Using closed-loop automation, IBN  continuously verifies and enforces these intents through three processes \cite{njah2024ai}: (\textit{i}) \textit{Intent refinement}, where user-defined intents are translated into maintainable and adaptable network policies; (\textit{ii}) \textit{Intent activation}, which leverages analytics and ML to detect and resolve policy conflicts; and
(\textit{iii}) \textit{Intent assurance}, which integrates with SDN controllers and orchestration systems, using telemetry for continuous monitoring and adjustment.

\begin{figure}[!t]
\centering
\includegraphics[width=.9\linewidth]{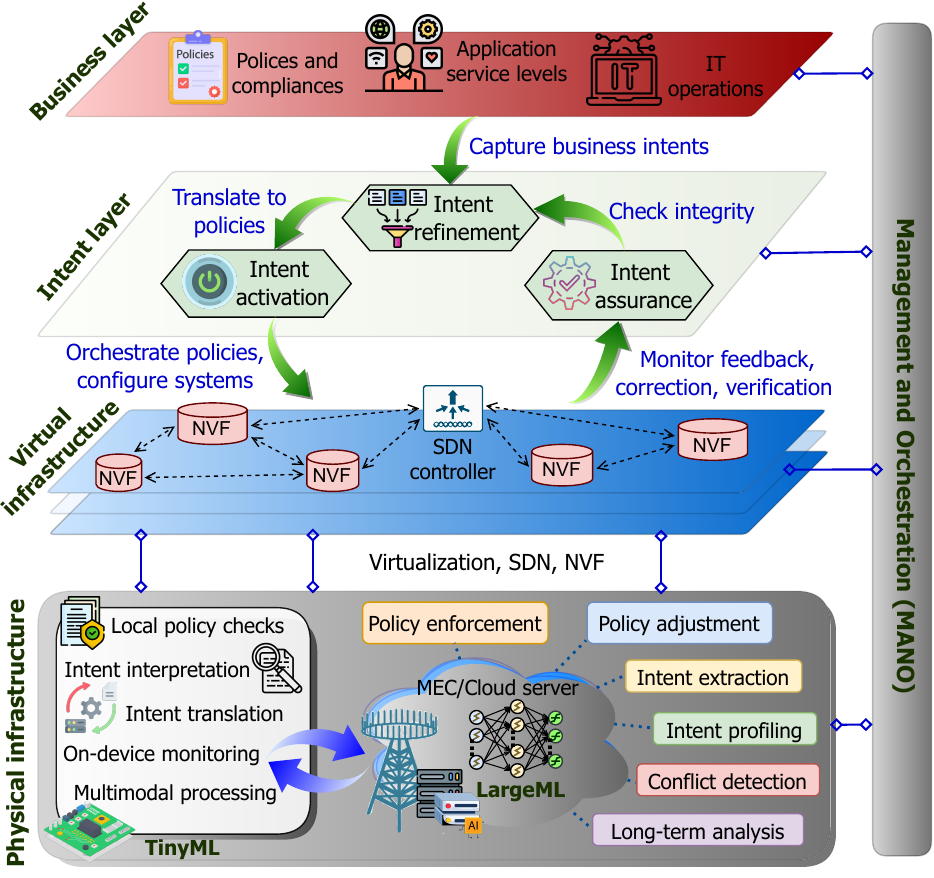}
\caption{IBNs with TinyML--LargeML integration.}
\label{Fig:App_IntentBased}
\end{figure}

The effectiveness of IBN depends on the accurate interpretation of user inputs (e.g., voice, text, multi-modal data). As shown in Fig.~\ref{Fig:App_IntentBased}, TinyML--LargeML integration supports this objective. 
TinyML provides lightweight, on-device AI for real-time intent detection and localized processing \cite{velasco2021end}, while LargeML aggregates intents across devices and translates them into network actions and policies using techniques such as TL, FTL, FFMs, SL, FSL, and FSFMs \cite{manias2024towards}.
A notable example is an intent-based on-device AI framework \cite{ahn2025intent}, where TinyML processes local sensor data (e.g., speech-to-text conversion and intent translation), while LargeML manages centralized policy coordination.
This reduces reliance on cloud-only processing, improving security, latency, and energy efficiency.
Another application is intent-based management for autonomous aerial vehicle swarm networks \cite{ajakwe2024time}, which combines TinyML with explainable AI, blockchain, and zero-trust architectures.
This system supports interpretable, tamper-proof intent predictions and detects false communications that may distort shared positional data.
Results in \cite{ajakwe2024time} demonstrate a 2.34\% accuracy gain, 2.97\% sensitivity improvement, latency reduced to 43.1 ms, and storage savings down to 0.0013 MB.

\subsection{Zero-touch Network}\label{SecV:ZTN}
A zero-touch network (ZTN) leverages AI, ML, and IBN to automate operations with minimal human intervention \cite{liyanage2022survey}. 
By handling tasks such as configuration, monitoring, and troubleshooting, ZTNs reduce operational costs, improve efficiency, and free administrators for strategic planning. 
They also enhance reliability by analyzing traffic and user requests in real time, adjusting parameters, resolving faults, and restoring services autonomously. Moreover, ZTNs align network behavior with organizational policies while adapting to growing demands and evolving requirements. Importantly, ZTNs integrated security mechanisms further enable real-time threat detection and protection.

\begin{figure}[!t]
\centering
\includegraphics[width=.85\linewidth]{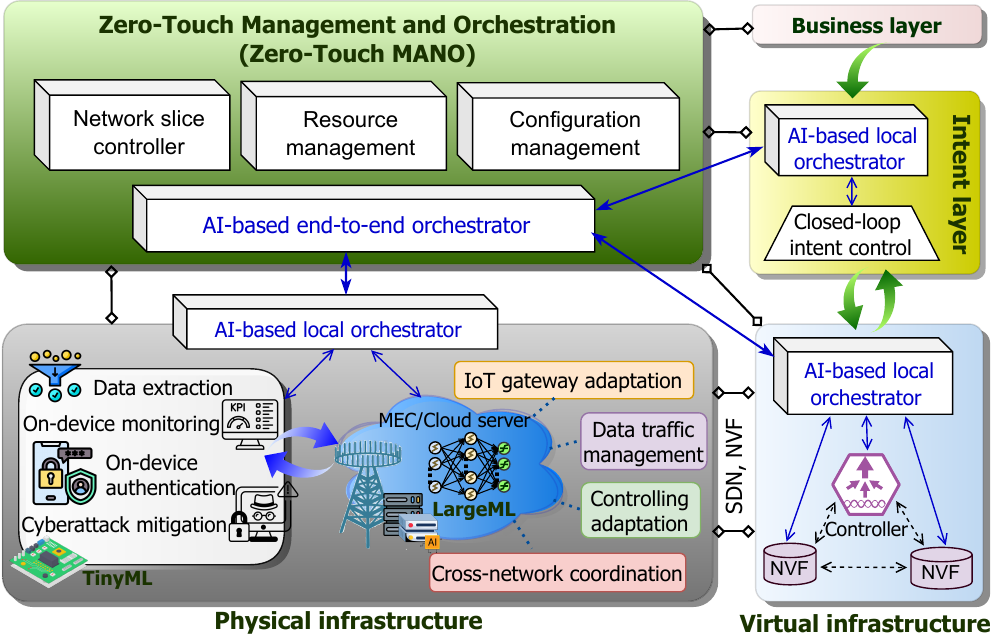}
\caption{ZTNs with TinyML--LargeML integration.}
\label{Fig:App_ZTN}
\end{figure}

Achieving zero-touch automation with net-zero emissions in AI-driven wireless networks demands energy-efficient TinyML and LargeML integration, low-power transceivers, lightweight AI models, and adaptive resource allocation \cite{abbas2023designing}. Recent research highlights a TinyML-LargeML ecosystem for 6G ZTNs (see Fig.~\ref{Fig:App_ZTN}).
In self-organizing networks, TinyML optimizes neural network training on IoT devices, while LargeML uses federated and deep learning at base stations or clouds to balance computational loads and minimize energy consumption \cite{shodamola2021towards}.
In mobile networking, TinyML allows autonomous IoT operations, while LargeML develops coverage maps and offers AI-as-a-service (AIaaS), as shown in NanoDeploy Automator \cite{samaras2024unlocking}. 
In industrial IoT, TinyML processes data locally, while LargeML enhances efficiency and traffic management at gateways and clouds, as seen in O-RAN-based ZTNs for smart factories \cite{wang2022deep}. 
Cross-layer cybersecurity frameworks, like SH-CASH, utilize AutoML to secure layers, with TinyML ensuring service-level agreement compliance in O-RAN architectures \cite{yang2025towards}.


\subsection{Brain-level Metaverse}\label{SecV:brain-level-metaverse}
%

\begin{figure}[!t]
\centering
\includegraphics[width=.8\linewidth]{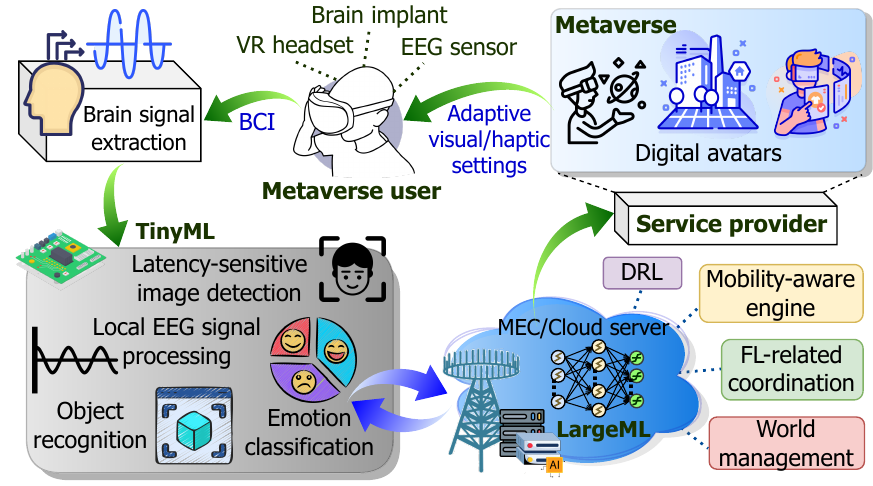}
\caption{BrainMeta with TinyML and LargeML.}
\label{Fig:App_BrainMeta}
\end{figure}

{The brain‑level metaverse (BrainMeta) links brain computer interfaces (BCIs) with immersive virtual worlds, allowing users to control avatars, interact with objects, and navigate environments through brain signals \cite{Zhu2024Apr}. BCIs range from non‑invasive EEG headsets \cite{Sterniuk2021Jan} to invasive implants \cite{Das2020Jan}, enabling personalized experiences in virtual learning, gaming, remote work, and neurorehabilitation while expanding accessibility for users with disabilities. Yet BrainMeta remains early‑stage and must overcome challenges in designing brain‑inspired cognitive architectures under strict size, weight, power, and processing limits \cite{yoo2021neural}, as well as the need for efficient, high‑performance models for neural signal interpretation, cognition support, and adaptive content generation, motivating integrated TinyML–LargeML deployment (see Fig.~\ref{Fig:App_BrainMeta}).

TinyML can leverage neuromorphic chips \cite{Kudithipudi2025Jan} to support diverse neural message formats, customize processing in brain‑inspired devices, localize memory, and accelerate tasks such as image recognition and classification, while edge‑based TinyML processors reduce latency and resource use through circuits like winner‑take‑all implementations on FPGA hardware \cite{khajooei2023super}. LargeML complements this with cloud‑edge‑end collaborative models, mobility‑aware pre‑rendering, and diffusion‑based adaptive rendering \cite{wang2025large}, enhancing data collection from TinyML devices and improving metaverse quality. Techniques include extracting generalized representations from large‑scale EEG datasets, segmenting signals into channel arrays, and applying vector‑quantized neural spectrum prediction for training neural analyzers \cite{jiang2024large}. Additionally, AMFL \cite{qiao2024amfl} combines FL and DRL to handle non‑IID data, reduce resource overhead, and improve QoE for resource‑constrained, human‑centric AR devices.}
\section{Challenges and Future Research Directions}
\label{Sec:challenges}

\subsection{Standardization\label{sec:standard}}
5G faces challenges with dynamic user demands and traffic, while 6G aims to enable sustainable applications like holography and the metaverse through TinyML–LargeML integration \cite{yang20226g}. This integration is designed for user-centric services that adapt to changing behaviors and networks. However, standardization issues impede the development of a fully integrated ecosystem. 
The ITU-R identifies RAN slicing as crucial for 6G virtual network sharing \cite{wp5d2022future}, requiring a unified O-RAN architecture that supports diverse air interfaces and AI functions. This architecture should offer customizable service parameters, but shared infrastructure raises concerns about transparency, reliability, and privacy.
Standardization bodies are advancing AI-driven wireless architectures. 3GPP's Releases 17 \cite{TR-37.817} and 18 \cite{TR-38.743} enhance 5G RAN with AI for intelligence and efficiency, while the O-RAN Alliance integrates AI for orchestration and control \cite{lin2023embracing}.

However, challenges remain in integrating TinyML and LargeML across the network lifecycle. Proposals include a secure TinyML edge AI architecture \cite{shabir2023toward}, explainable AI for decision transparency \cite{wang2021applications}, and a testbed for trustworthy network management \cite{rezazadeh2024toward}. Sustainability efforts are supported by an energy cost-of-AI metric \cite{chou2024towards}.
At a higher level, a shift to a multi-tier mNode model is proposed \cite{yang20226g}, featuring distributed nodes and network AI logic. Additionally, a digital twin framework for AI-RAN can manage digital twins and enable real-time adaptability in dense networks \cite{faye2024integrating}.

\subsection{Resource Management and Orchestration}

{6G networks face rising orchestration and resource management pressures due to wireless channel uncertainty, dense ISAC deployments, and carbon‑reduction requirements. AI/ML improves interference control, energy efficiency, and reliability \cite{du2024enhancing}, but also raises concerns about privacy, computational load, and scalability. New directions include DRL paired with diffusion models for adaptive optimization \cite{du2024enhancing} and FTL with blockchain‑based incentives for low‑latency, privacy‑preserving coordination \cite{wang2023incentive}. TinyML remains constrained by compute trade‑offs, vanishing gradients that demand hardware acceleration, memory limits for semantic communication, limited real‑world data, and environmental variability \cite{tong2022nine}. LargeML, despite distributed learning, still struggles with data segmentation, transmission errors in FL, and strict privacy and coordination needs. Future methods such as model‑agnostic meta‑learning \cite{finn2017model} and accretionary learning \cite{wei2023accretionary} may help alleviate these challenges.

At the orchestration layer, integrating AI into 6G network slicing blends NFV‑based virtualization with SDN's centralized control: a global SDN controller handles planning and provisioning, while AI‑enhanced local controllers optimize slice‑level performance. Cross‑domain AI coordination introduces challenges in data and knowledge sharing, training alignment, inference synchronization, and resource allocation \cite{li2022distributed}. As 6G shifts from centralized 5G models to distributed learning, learning-model orchestration becomes more complex and depends on node capacity, data properties, resource availability, model size, latency, and network dynamics. Assigning specialized roles across slices enables scalable, efficient learning in 6G environments, e.g., horizontal FL for decentralized data, hierarchical FL for reduced communication, TL for efficient reuse, vertical FL for cross‑domain learning, SL for heterogeneous resources, multi-agent RL for local coordination, and swarm learning for secure joint training.}


\subsection{Advanced Security and Privacy-Preserving Techniques}

{As TinyML–LargeML integration becomes central to next‑generation wireless networks, security and privacy concerns intensify, especially around adversarial threats, data protection, model integrity, and user privacy. TinyML's limited edge‑device resources make sensors, smartphones in FL, and wearables vulnerable to attacks, but adversarial training strengthens robustness \cite{kuhnel2024semi}, homomorphic encryption protects data confidentiality and integrity \cite{ullah2024homomorphic}, differential privacy obscures sensitive information \cite{aouedi2024survey}, and advanced encryption secures model parameters during transmission \cite{ajagbe2024advanced}. LargeML, operating across broader and multi‑stakeholder environments, also requires strong safeguards: model distillation reduces computational demands while preserving performance \cite{yang2023categories}, GAN‑based simulations improve cyber‑resilience \cite{vu2024applications}, and blockchain ensures model integrity by hashing and tracking TinyML model versions \cite{aouedi2024survey}. Additional privacy‑preserving techniques, including data shuffling, machine unlearning \cite{nguyen2026towards}, synthetic data generation, private teacher‑ensemble aggregation, Rényi differential privacy, secure cross‑silo FL, and decentralized knowledge distillation, further reinforce LargeML deployments.}

\subsection{Real-time, Lightweight Intelligence}

{Beyond‑5G and 6G networks are shifting from traditional M2M automation toward human‑centric architectures, demanding AI systems that provide real‑time, lightweight intelligence while meeting user‑experience and sustainability goals. Joint TinyML–LargeML deployment expands these capabilities but also introduces challenges. TinyML devices span a wide range of protocols, latency profiles, and hardware, from Arduino Nano (64 MHz, 1 MB flash, 256 KB RAM) to Raspberry Pi (1.8 GHz, 32 GB flash, 8 GB RAM) \cite{abadade2023tinyml,Lin2023Oct}, emphasizing the need for compact, unified frameworks that support plug‑and‑play deployment, efficient processing, and lightweight runtimes \cite{huang2024plug}. LargeML, by contrast, handles massive multimodal inputs and depends on prompt engineering, model partitioning, and hyperparameter tuning, all of which incur high training costs, power consumption, and memory demands \cite{raiaan2024review}. Emerging techniques such as model‑adaptive prompts \cite{chen2024mapo} and pluggable lightweight tuning \cite{chen2021lightner} can improve explainability and accelerate inference without disrupting life‑cycle pipelines.

Integration remains limited by the lack of standardized platforms and tools for workflow management, data coordination, and runtime compatibility. Large models often require specialized libraries, compilers, and runtimes with inconsistent input/output specifications \cite{wu2025consolidating}, making iterative refinement necessary for reliable outputs. As a result, progress depends not only on optimizing TinyML and LargeML individually but also on developing guidelines and evaluation frameworks that define task‑specific roles, interoperability requirements, and life‑cycle management strategies across domains \cite{he2024ultraeval}.}

\subsection{AI‑enabled 6G Communication-Sensing-Computing}
\label{Sect:VI-E}

{
Integrating TinyML's real‑time, low‑power intelligence with LargeML's high‑capacity reasoning creates a unified AI layer for 6G, where communication, sensing, and computation operate as a single adaptive system \cite{11062661}. In this hierarchy, TinyML provides fast on‑device perception through lightweight feature extraction and inference, while LargeML offers deeper interpretation and coordination through global optimization, multimodal fusion, and long‑horizon prediction. To do so, TinyML first semantically compresses sensed data for efficient transmission \cite{11103469}, LargeML then uses these representations to optimize routing, beamforming, and spectrum allocation, and finally, LargeML's predictions guide TinyML nodes to adjust sensing rates or modalities, forming adaptive networks responsive to mobility, interference, and environmental changes. This synergy supports scalable multimodal sensing, energy‑efficient operation, and human‑centric services  \cite{9687468}, where edge devices capture fine‑grained context and cloud models infer intent or anticipate user needs. 

However, integrating TinyML and LargeML in 6G introduces significant challenges. First, model partitioning must balance latency, energy, and accuracy across heterogeneous devices \cite{11103469}. Second, synchronizing TinyML's millisecond‑scale inference with LargeML's slower global reasoning requires new timing and scheduling mechanisms. Third, interoperability is hindered by diverse hardware, runtime environments, and data formats across edge devices \cite{11098465}. Fourth, security and privacy risks increase as sensitive edge data interacts with large‑scale models, demanding robust protection against adversarial attacks, leakage, and model manipulation. Finally, additional 6G‑specific constraints (i.e., URLLC requirements, dynamic spectrum conditions, mobility‑induced channel variation, and sustainability targets  \cite{11062661,9687468}) further complicate joint deployment.

}

\subsection{AI-Native and Collective AI for 6G}

{AI‑native systems embed intelligence at the core of their operation, aiming to solve complex tasks through data‑driven reasoning, adapt to dynamic environments, fuse foundational knowledge with new insights, and maintain transparency and trustworthiness for resilient autonomous networks \cite{iovene2023defining}. Integrating TinyML and LargeML into these architectures introduces challenges such as coordinating constrained TinyML with powerful LargeML for zero‑touch automation, defining their roles across heterogeneous infrastructures, ensuring interoperable lifelong learning, managing AI life cycles for fairness and reliability, and balancing joint performance without degrading TinyML accuracy. Collective AI extends this vision through multi‑agent collaboration, where agents learn locally toward shared goals without direct communication \cite{zheng2018magent}; platform‑level collective reinforcement learning accelerates convergence and reduces redundant training \cite{li2022intelligent}, while system‑level swarm intelligence supports IoT applications in dynamic environments \cite{rosenberg2024towards}, though it still lacks unified reasoning and coordination frameworks.

Integrating TinyML and LargeML into collective AI creates new opportunities but also raises several challenges: sustaining agent participation through incentive mechanisms, ensuring efficient spectrum use in dense 6G IoT deployments, and addressing security, trust, and privacy concerns as shared outputs may expose sensitive information while blockchain remains resource‑intensive for constrained devices. Human-AI collaboration further requires cognitive architectures capable of interpreting social signals, and ethical considerations, including transparency, bias mitigation, and accountability, are essential, especially as errors in LargeML outputs can have significant real‑world consequences.}

\subsection{Application Prospects and Outlook for 6G and AI}
\label{Sect:VI-G}
{ 
6G will unify TinyML at the edge with large-scale AI models deployed across the network and cloud, creating an intelligent and adaptive wireless environment. This integration will enable real-time sensing, IBN, semantic communication, and automated SDN/NFV orchestration, thereby forming the foundation for immersive XR, holographic telepresence, and large-scale digital twins
\cite{huawei20226g,6G_AI}. Early scenarios already reflect this direction: in smart factories, TinyML sensors detect anomalies locally and send only semantic insights, while edge AI coordinates predictive maintenance and robotic actions with sub‑second latency. At the city scale, distributed TinyML nodes monitor environmental and public‑safety conditions and feed compact features into collective AI systems that coordinate drones, traffic flows, and emergency responses through privacy‑preserving FL.

Over the longer term, 6G is expected to evolve into an AI‑native network where intelligence is embedded across all layers, and devices collaborate through distributed learning and collective decision making \cite{6G_AI}. This will enable ZTN operation, self‑healing infrastructures, and brain‑level Metaverse experiences that tightly integrate communication, sensing, and computation. Realizing this vision requires advances in O‑RAN standardization, trustworthy and explainable AI, secure FL, and efficient orchestration of communication, computing, and energy resources. As these foundations mature, the fusion of 6G and AI will support resilient industrial automation, new service models, and globally scalable intelligence.
}



\section{Concluding Remarks}
\label{sec:Conclusion}
{TinyML--LargeML integration is set to reshape AI design for 6G by enabling efficient, intelligent solutions to challenges in resource allocation, network management, security, and privacy. This survey reviews the foundations of this convergence, covering design principles, operational mechanisms, and performance evaluation while outlining the motivations and technical requirements driving bidirectional integration and its emerging application scenarios across 6G ecosystems. This survey also identifies key open challenges and promising research directions, including O‑RAN standardization, cross‑layer resource orchestration, advanced security and privacy‑preserving methods, real‑time lightweight AI, AI-enabled 6G communication-sensing-computing, the shift toward AI‑native and collective AI architectures, and application prospects and outlooks for 6G and AI. Continued progress in these areas is essential, and the insights provided here aim to guide and support future  6G‑enabled intelligent systems.}

\balance
\bibliographystyle{IEEEtran}
\bibliography{Final_Reference.bib}

\end{document}